\documentclass[aps,preprint,nofootinbib]{revtex4}
\usepackage{sublabel}
\usepackage{graphicx}
\usepackage{epsf}
\usepackage{wrapfig}
\usepackage{epsfig}
\def\Vec#1{\mbox{\boldmath $#1$}}
\def\bra#1{\langle#1|}
\def\ket#1{|#1\rangle}

\def\beq{\begin{equation}}
\def\eeq{\end{equation}}
\def\nn{\nonumber}
\def\beqy{\begin{eqnarray}}
\def\eeqy{\end{eqnarray}}
\def\ie{\textit{i.e.}}
\def\eg{\textit{e.g.}}
\def\mo{\mathcal{O}}
\def\mh{\hat{\mathcal{O}}}

\newcommand{\ossi}{$^{16}O$}

\newcommand{\vuotto}{$V8^\prime$}

\begin{document}
\vskip 2mm \date{\today}\vskip 2mm
\title{Ground-state energies, densities and momentum distributions in closed-shell
 nuclei
  calculated within a cluster expansion approach and realistic interactions}
\author{M. Alvioli}
\author{C. Ciofi degli Atti}
\address{Department of Physics, University of Perugia and
      Istituto Nazionale di Fisica Nucleare, Sezione di Perugia,
      Via A. Pascoli, I-06123, Perugia, Italy}
\author{H. Morita}
\address{ Sapporo Gakuin University, Bunkyo-dai 11, Ebetsu 069-8555,
Hokkaido, Japan}
\vskip 1cm
\begin{abstract}
\noindent
A linked cluster expansion suitable for the treatment  of ground-state
properties
of complex nuclei, as well as of various particle-nucleus scattering processes,
 has been used to calculate
the ground-state energy, density and momentum distribution of $^{16}O$
and $^{40}Ca$ using  realistic interactions.
First of all,  a  benchmark calculation for the ground-state energy 
has been  performed  using the truncated $V8^\prime$ potential,  and
consisting in the comparison
of our results with the ones obtained by the Fermi Hypernetted Chain
approach, adopting  in both cases
the same mean field wave functions and the same correlation functions.
 The results exhibited a nice
agreement between the two methods. Therefore, the approach has been 
applied to the calculation of the ground-state energy, density and momentum distributions
of $^{16}O$
and $^{40}Ca$ using the full $V8^\prime$ potential, finding again
a satisfactory agreement with the results based on  more advanced approaches 
where higher order cluster contributions are taken into account.
It appears therefore
that the cluster expansion approach can  provide accurate approximations
for various diagonal and non diagonal density matrices, so that
it could be used for a reliable evaluation of nuclear effects
in  various  medium and high energy scattering processes off nuclear targets.
The developed approach can be readily
generalized
to the treatment of  Glauber type final state interaction effects in inclusive,
semi-inclusive
and exclusive processes off nuclei at medium and high energies.
\end{abstract}
\maketitle
%\pacs{PACS}

%%%%%%%%%%%%%%%%%%%%%%%%%%%%%%%%%%%%%%%%%%%%%%%%%%%%%%%%%%%%%%%%%%%%%%%%%%%

\section{Introduction}

\noindent
The knowledge of the nuclear wave function, in particular its
 most interesting and unknown  part, {\it viz}
the correlated one, which is  predicted by  realistic many-body calculations
   to strongly deviate from a mean field description,
is not only
a prerequisite for understanding the details of bound hadronic systems, but
is becoming at present a necessary condition for a correct description of medium
and high energy scattering processes off nuclear targets; these, in fact,
represent nowadays an efficient tool for the investigation
of several high energy problems, e.g. color transparency,  hadronization, the
properties of  dense
 hadronic matter, etc.
which manifest themselves only in the nuclear medium.
The necessity of an accurate treatment of the effects of the medium
in high energy scattering  process is becoming a relevant issue in
hadronic physics. The problem is not trivial, for one has first to solve
 the many body problem
and then has to find a way to  describe  scattering processes in terms of realistic
many-body wave functions.
The  difficulty mainly arises because even if  a reliable and manageable
  many-body description of the ground state is developed, the
problem remains of the calculation of the final state.
 In the case of
\textit{few-body systems}, a consistent treatment of initial state correlations (ISC)
 and final
state interaction (FSI) is nowadays possible at low energies by
solving the Schr\"odinger equation for the bound and continuum
states (see \eg \cite{gloeckle,pisa,he3bench} and References therein
quoted), but at high energies,  when the number of partial waves
sharply increases  and  nucleon excitations can occur, the
Schr\"odinger approach  becomes impractical and other methods have
to be employed. Moreover, in the case of \textit{complex nuclei},
additional difficulties arise due to the approximations which are
still necessary to solve the many-body problem. As a matter of fact,
in spite of the relevant progress made in recent years in the
calculation of the properties of light nuclei (see $e.g.$\,
\cite{wiri,pie01,vary,fab01,fab02}), much remains  to be done,
also in view that the results of very sophisticated calculations
($e.g.$\,\, the variational Monte Carlo ones \cite{pie01}), show
that the wave function which minimizes the expectation value of the
Hamiltonian provides a very poor nuclear density; moreover, the
structure of the best trial wave function is so complicated, that
its use in the calculation of various processes at intermediate and
high energies appears to be not easy task. It is for this reason
that the evaluation of  nuclear effects in medium and high energy
scattering processes is usually  carried out within simplified
models of nuclear structure. As a matter of fact, ISC  are often
introduced by a procedure which has little to recommend itself,
namely the expectation value of the transition operator is evaluated
with shell model (SM) uncorrelated wave functions  and the initial two-body
wave function describing the independent relative motion of two nucleons
is replaced by a
phenomenological correlated wave function. Recently, however,
important progress
 has been
made, in that ISC have been introduced from the beginning using
 correlated wave functions and
cluster expansion techniques. Central Jastrow-type  correlations have been often used
to investigate the effects of ISC on various scattering
processes off complex nuclei induced by medium energy leptons like, e.g. $A(e,e')X$ \cite{SRCinclusive},  $A(e,e'p)X$
\cite{SRCexclusive} and
$A(e,e'2N)X$ \cite{SRCtwonucleon} processes. Calculations of inclusive electron scattering
has also been performed using realistic many body wave functions and spectral functions
within various
approximations \cite{omar,EMC}, and  non central correlations
have  been recently introduced in the calculation of $A(e,e'p)B$ and $A(e,e'2N)B$
 \cite{11a,11b}. In spite of this progress, further work remains to be done
 to achieve a full  consistent treatment of
both ISC and FSI in intermediate and high energy scattering off complex nuclei.
This would be particularly
urgent, as far as
various high energy phenomena are concerned,
e.g.  exclusive processes at high momentum transfer
\cite{SRCtransparency},
inclusive \cite{SRCcolor} and semi inclusive \cite{SRCDIS}
 hadron production in Deep Inelastic
Scattering, and others, which might also
require a careful treatment of nuclear effects. As a matter of fact, a recent
calculation \cite{clada} of  the integrated nuclear transparency in the processes
 $^{16}O(e,e'p)X$ and $^{40}Ca(e,e'p)X$
 performed within a cluster expansion approach, including realistic
central and  tensor correlations,  shows
that the results do depend both on the SM and ISC parameters,
which have therefore to be fixed from firm criteria, e.g. from the
calculation of the static properties of nuclei, like the ground-state
energy and the density distribution.

For such a reason, we have undertaken the calculation of the basic ground-state
properties (energies, densities and momentum distributions)
 of complex nuclei within
a framework which can  easily be generalized to the treatment of various scattering
processes, keeping  the basic features of ISC as predicted by the structure of
realistic Nucleon-Nucleon (NN) interactions.
Our approach is presented in detail in this paper,  which is organized as follows:
in Section II some basic ideas concerning  the application  of the
cluster expansion techniques to
the approximate
solution of the nuclear many-body problem are
recalled; the cluster expansion  used in the calculations is  described in Section
III; the ground-state energy calculations for $^{16}O$ and $^{40}Ca$ are
 presented in Section
IV, where the results a benchmark calculation aimed at a comparison
of our results  with the results obtained within
the Fermion Hypernetted Chain (FHNC) approach \cite{fab01,fab02,30a} is presented;
  the results of the calculations of  the charge densities and momentum distributions are given
in Sections V and VI, respectively; the diagrammatic representation of the latter
quantities within the cluster-expansion approach are illustrated 
in Section  VII; the Summary
 and Conclusions are presented in Section VIII.

Preliminary results of our calculations have been presented in Ref.
\cite{alv01,alv02}.

\section{The Correlated wave functions}

It is well known that
if nuclei are considered to be aggregates of point-like nucleons with
the same properties and interactions as the free ones, and, moreover,
all degrees of freedom but nucleonic ones are frozen, the nuclear many-body
problem reduces to the search of the eigenvalues and eigenfunctions of
the following Schr\"odinger equation \cite{pri01}
\beq
\label{dueuno}
\left[\,\sum^A_{i=1}\,\frac{\Vec{p}^2_i}{2\,M_N}\,+\,\hat{V}_{eff}(1,2,...,A)\right]
        \,\psi_n(1,2,...,A)\,=\,E_n\,\psi_n(1,2,...,A)
\eeq
where $M_N$ is the nucleon mass and the effective interaction includes
many-body interactions between the costituents, \ie
\beq
\label{duedue}
\hat{V}_{eff}(1,2,...,A)\,=\,\sum_{i<j}\hat{v}_2(i,j)\,+\,\sum_{i<j<k}
        \hat{v}_3(i,j,k)\,+\,...\,+\,v_A(1,2,...,A)\,.
\eeq
Within the so called \textit{standard model} of nuclei \cite{cio01},
which will be considered from now on,
many-body interactions are disregarded and Eq. (\ref{dueuno})
is solved keeping only the two-body interaction $\hat{v}_2(i,j)$, whose
form is determined from two-body bound and scattering data. We will consider
therefore the following nuclear Hamiltonian:
\beq
\label{hamilt}
\hat{H}\,=\,\hat{T}\,+\,\hat{V}\,
        =\,-\frac{\hbar^2}{2\,M_N}\,\sum^A_{i=1}\,\nabla^2_i\,
                \,+\,\sum_{i<j}\,\hat{v}_2(\Vec{x}_i,\Vec{x}_j)\;,
\eeq
where the vector $\Vec{x}$ denotes the set of nucleonic degrees of freedom,
\ie{ } $\Vec{x}\equiv(\Vec{r};\Vec{\sigma};\Vec{\tau})$, with $\Vec{r}$, $\Vec{\sigma}$
and $\Vec{\tau}$ denoting the spatial, spin and isospin coordinates,
respectively.
We will try to find the solution of the Schr\"odinger equation pertaining to
the ground-state of the nucleus, \ie:
\beq
\label{accapsi}
\hat{H}\,\psi_o\,=\,E_o\,\psi_o
\eeq
and to this end we will look for the ground-state wave function (WF) $\psi_0$
which minimizes   the expectation value of the Hamiltonian
\beq
\label{ezero}
\langle\,\hat{H}\,\rangle\,=\,\frac{\bra{\psi_o}\,\hat{H}\,\ket{\psi_o}}{\langle
  \psi_o\,|\,\psi_o \rangle}\,\ge\,E_o\,.
\eeq
As trial WF we will use a  correlated WF of the
following form \cite{clark}
\beq
\label{psi1}
\psi_o(\Vec{x}_1,...,\Vec{x}_A)\,=\,\hat{F}(\Vec{x}_1,...,\Vec{x}_A)\,
        \phi_o(\Vec{x}_1,...,\Vec{x}_A)\,,
\eeq
where $\phi_o$  is a SM ,  mean-field WF describing
the independent particle motion, and $\hat{F}$ is  a symmetrized \textit{correlation
operator}, which generates correlations into the mean field WF;
the correct symmetry of the WF is guaranteed by $\phi_o$.

As in any variational approach, the central problem here is to give
an explicit form to the trial WF (Eq. (\ref{psi1})); whereas for $\phi_o$ any realistic
SM  WF can be considered a physically sound approximation,
the choice of the form of the operator $\hat{F}$ is not clear \textit{apriori}.
However one can be guided by the knowledge of the basic features of
the force acting between the considered hadrons. Nowadays the
nucleon-nucleon interaction can be cast in the following form
(\cite{twonucl}):
\beq
\label{defpotere}
\hat{V}\,=\,\sum_{n=1}^{N}\,v^{(n)}(r_{ij})\,\mh^{(n)}_{ij}\,,
\eeq
where $r_{ij}=|\Vec{r}_i-\Vec{r}_j|$ is the relative distance of
nucleons $i$ and $j$,  and $n$, ranging up to $N=18$, labels
the state-dependent operator $\mh^{(n)}_{ij}$:
\beq
\label{newuno}
\mh^{(n)}_{ij}\,=\,\left[1\,,\,\Vec{\sigma}_i\cdot\Vec{\sigma}_j\,
        ,\,\hat{S}_{ij}\,,\,\left(\Vec{S}\cdot\Vec{L}\right)_{ij}\,,
        \,L^2\,,\,L^2\Vec{\sigma}_i\cdot\Vec{\sigma}_j\,,\,
      \left(\Vec{S}\cdot\Vec{L}\right)^2_{ij}\,,..
        \right]\,\otimes\,\left[1\,,\,\Vec{\tau}_i\cdot\Vec{\tau}_j\right]
\eeq
Accordingly, the operator $\hat{F}$ is written as
\beq
\label{newdue}
\hat{F}(\Vec{x}_1,\Vec{x}_2\,...\,\Vec{x}_A)\,=\,\hat{S}\,\prod^A_{i<j}\,\hat{f}(r_{ij})
\eeq
with
\beq
\hat{f}(r_{ij})=\sum_{n=1}^{N}\,\hat{f}^{(n)}(r_{ij})\;\;\hspace{2cm}
        \hat{f}^{(n)}(r_{ij})=f^{(n)}(r_{ij})\,\hat{O}^{(n)}_{ij}\,.
\label{corrop1}
\eeq

The variational principle requires the full evaluation of Eq.
(\ref{ezero}) which, obviously, is no easy task due to the
structure of $\psi_0$. We will evaluate the expectation value of the
Hamiltonian (\ref{ezero}), using the cluster expansion techniques
\cite{clark}, adopting a specific cluster expansion to be described
in the next sections.

%%%%%%%%%%%%%%%%%%%%%%%%%%%%%%%%%%%%%%%%%%%%%%%%%%%%%%%%%%%%%%%%%%%%%%%%%%%%%%%%%%%%%%%%%%%%%%%%%%%%%

\section{The cluster expansion}

The evaluation of the expectation value of $\hat{H}$ is object of intensive
activity which in the last few years has produced considerable results: the
approximate solution of the Schr\"odinger equation by means of Monte Carlo
methods, for example, has reached a great level of accuracy, and the ground-state
properties of nuclei with $A=16$ have been obtained with a full evaluation
of Eq. (\ref{ezero}) \cite{wiri,pie01}; exhaustive calculations have also been
performed within the FHNC approximation \cite{fab01,fab02,30a}.
Nevertheless, the level of complexity of these calculations often prevents the WF
to be used with reasonable ease in other nuclear-related problems, such
as nuclear reactions. Our goal is to present a more economical,   but effective
method for the calculation of the expectation value of any quantum
mechanical operator $\mh$ in the many-body ground-state described by the WF
$\psi_o$, \ie:
\beq
\label{omedio1}
\langle\mh\rangle\,=\,\frac{\bra{\psi_o}\mh
        \ket{\psi_o}}{\bra{\psi_o}\psi_o\rangle}\,;
\eeq with $\psi_o$ having the structure of Eq. (\ref{psi1}). In the
present section we are going to introduce a cluster expansion
technique in order to evaluate Eq. (\ref{omedio1}). To begin with, a
generic operator $\mh$ in Eq. (\ref{omedio1}) will be considered in
the following Section, while in the next Section, it will
be specialized to the Hamiltonian and to the one- and two-body
density operators.

Various types of cluster expansions have been used in the past
to calculate the ground-state properties of
nuclei  (see e.g. \cite{rafa,rafa1}); in these calculations
, mainly aimed at investigating the convergence of the expansion,
 simple models of the NN interaction have been usually used
. In this paper we use an expansion which has never been used previously to calculate gropund-
state properties of nuclei in terms of realistic interactions.
 The expansion we are going to use has been originally developed in Ref.
\cite{gau01} (see also \cite{boh01} and \cite{dal01}); the main feature
of such an expansion is that it is linked and number conserving.
The latter property means that the normalization of any observable is
provided by the normalization of the mean field WF, \ie, by
the first term of the expansion: the contribution of all other terms to the
normalization vanishes analytically order by order. The expansion, to be called
the  \textit{$\eta$-expansion,} has been originally used to obtain the lowest
order contribution to the diagonal one-body density (OBD) matrix,
 $\hat{\rho}^{(1)}(\Vec{r}_1)$ \cite{gau01}, and to the one-body mixed density (OBMD)
 matrix, $\hat{\rho}^{(1)}(\Vec{r}_1,\Vec{r}^\prime_1)$ \cite{boh01},
 using central correlations only (\ie, Eqs.
 (\ref{newdue}) and (\ref{corrop1}) with $N=1$). Subsequently
\cite{ben01}, the  lowest order expansion of the OBMD operator has
been generalized to take into account also the non central
spin-isospin and tensor-isospin  correlations $f^{4}$ and $f^{6}$ in
Eq. (\ref{corrop1}), which turned out to be the most relevant non
central correlation functions in Nuclear Matter (\cite{nmf3}), as
well as all correlations up to $N=6$  \cite{fab01}\footnote{ The
approximation which includes only the components $n = \{1,4,6\}$ is
usually referred to as the \textit{$f_3$ approximation},  whereas
the approximation which includes all correlations up to $N = 6$ is
referred to as the \textit{$f_6$ approximation}.}. During the last
few years, lowest order expansions have also been applied, within
the central correlation approximation, to the calculation of the
two-body density matrix \cite{dim01} and of various transition
matrix elements appearing in inclusive, $A(e,e^\prime)X$, and
exclusive, $A(e,e^\prime p)B$ and $A(e,e^\prime 2 N)B$, processes
(see \textit{e.g.} Refs.
\cite{SRCinclusive,SRCexclusive,SRCtwonucleon}). As already
mentioned, the expansion has also been used, within the $f_3$
approximation, to calculate the nuclear transparency in the
semi-inclusive process $A(e,e^\prime p)X$ \cite{clada}. To our
knowledge,  the $\eta$-expansion has never been used to calculate
the ground-state energy of complex nuclei with a realistic
interaction. It is precisely the central aim of our work to present
a detailed report of the results of the calculation of the
ground-state energy, density and momentum distributions of complex
nuclei using the $\eta$-expansion and realistic interactions.

Let us first of all recall the basic features of the expansion.
Following the formal expression for $\psi_o$ in \ref{psi1}, and taking
the correlation operator $\hat{F}$ as in \ref{corrop1}, one writes
\beqy
\hat{F}^2&=&\prod_{i<j}\,\hat{f}^2(r_{ij})\,=\prod_{i<j}\,
        \left(1+\hat{\eta}(r_{ij})\right)\,=\nn\\
        &=&1+\sum_{i<j}\hat{\eta}_{ij}\,+\,\sum_{(ij)<(kl)}\hat{\eta}_{ij}
                \,\hat{\eta}_{kl}\,\,+\,...
\label{corrop2}
\eeqy
where
\beq
\hat{\eta}_{ij}\,\equiv\,\hat{f}^2_{ij}\,-\,1
\eeq
and $\langle|\eta|^2\rangle$ will play the role of a \textit{small} expansion
parameter, in that its
expectation value on the reference state $\phi_o$ is small.
We use the notation $\eta_{ij}\equiv\eta(r_{ij})$.
In what follows, when dealing with state-dependent operators,
we have to bear in mind that they do not commute with each other;
moreover, since the same operators appear both in the potential and in the
WF, it is worth defining the following quantities:
\begin{itemize}
\item[1)]{$\label{notation1}
\langle A\rangle\,\equiv\,\bra{\phi_o}\,A\,\ket{\phi_o}\,,$
where $A$ is an arbitrary quantity;}
\item[2)]{$\hat{\eta}_{ij}\,\mh\,\equiv\,\hat{f}_{ij}\,
        \mh\,\hat{f}_{ij}\,-\,\mh\,;$}
\item[3)]{$\hat{\eta}_{ij}\,\hat{\eta}_{kl}\,\mh\,\equiv\,
  \hat{f}_{ij}\,\hat{f}_{kl}\,\mh\,\hat{f}_{ij}
  \,\hat{f}_{kl}\,-\,\hat{f}_{ij}\,\mh\,\hat{f}_{ij}\,-\,\hat{f}_{kl}\,\mh\,
  \hat{f}_{kl}\,+\,\mh\,;$}
\end{itemize}
and so on, where $\hat{O}$ is the operator appearing in Eq. (\ref{omedio1}).
Let us now perform the expansion of the expectation value (\ref{omedio1}).
Keeping in mind the described recipes, the quantity  $\hat{F}^\dagger\hat{F}$
 is expanded
both in the numerator and the denominator and  all terms containing
the same number of functions $\eta_{ij}=\eta(ij)$ are collected, obtaining
\beq
\label{omedio2}
\langle\mh\rangle\,=\,\mo_0\,
        +\,\mo_1\,+\,\mo_2\,+\,...\,+\,\mo_n\,+\,...\,+\,\mo_A\,,
\eeq

\noindent At $2$-nd order in $\eta$, one has, explicitly,
\beqy\sublabon{equation}
\mo_0&\equiv&\langle\mh\rangle\,,\label{eta1a}\\
\nn\\
\mo_1&=&\langle\sum_{ij}\hat{\eta}_{ij}\,\mh\rangle\,
        -\,\mo_0\,\langle\sum_{ij}\,
        \hat{\eta}_{ij}\rangle\,,\label{eta1b}\\
\nn\\
\mo_2&=&\langle\sum_{ij<kl}\hat{\eta}_{ij}\,\hat{\eta}_{kl}\,
        \mh\rangle\,
        -\,\langle\sum_{ij}\hat{\eta}_{ij}\,\mh\rangle\,
        \langle\sum_{ij}\hat{\eta}_{ij}\rangle\,+\label{eta1c}\\
        & &\,-\,\mo_0\,\left(\langle\sum_{ij<kl}\hat{\eta}_{ij}\,
        \hat{\eta}_{kl}\,\rangle\,-\langle\sum_{ij}\hat{\eta}_{ij}\rangle^2
        \right)\,;\label{eta1d}
\eeqy\sublaboff{equation}
where the term of order \textit{n} contains $\hat{\eta}$ ($\hat{f}$)
up to the \textit{n}-th (\textit{2n}-th) power.

Analyzing the structure of Eqs. (\ref{eta1d}) one realizes that there are
terms which are due to the expansion of the denominator
\beq
\frac{1}{1-x}\,\simeq \,1 \,+\, x \,+\, ...\,,
\eeq
e.g. the $2$-nd term in Eq. (\ref{eta1b}).

A nice feature of the $\eta$-expansion shows up at this point. Each
of the terms in the residual formul\ae{ }presents some
\textit{linked} and \textit{unlinked} contributions. What we mean by
this is quite self-explaining when expressed in terms of standard
Mayer diagrams \cite{mar01}, according to which in each linked term
the involved $n<A$ particle coordinates are connected either by an
$f$ factor, or by Pauli correlations (see Section VII). It is
precisely the expansion of the denominator which ensures that only
linked terms contribute to the overall expectation value, all
unlinked terms cancelling out amongst themselves. This feature turns
out to be very convenient from a computational point of view, for it
reduces the number of involved linked terms and allows one to obtain
a very systematic and general procedure. Eventually it should be
stressed that because of the non-commutative nature of the operators
$\mh^{(n)}_{ij}$ involving at least one common particle index, sets
of diagrams involving more than two particles appears in the
expectation value of the Hamiltonian expression within the cluster
expansion, already at first order.

\section{Application of the $\eta$-expansion to the nuclei $^{16}O$ and $^{40}Ca$: general
  formul\ae{ }and a benchmark calculation with truncated $V8^\prime$
  and $U14$ interactions}

\subsection{General formul\ae{ }in terms of density distributions}\label{sectionIVA}

Given the two-body interaction as in Eq. (\ref{defpotere}), the expectation value
of the Hamiltonian can be written in the following way \cite{mar01}:
\beq\label{hmatrix}
E_o\,=\,-\,\frac{\hbar^2}{2\,M_N}\,\int d\Vec{r}_1\left[\nabla^2\,
        \rho^{(1)}(\Vec{r}_1,\Vec{r}_1^\prime)\right]_{\Vec{r}_1=\Vec{r}_1^\prime}
        \,+\,\sum_n\;\int\;d\Vec{r}_1 d
        \Vec{r}_2\;v^{(n)}(r_{12})\rho^{(2)}_{(n)}(\Vec{r}_1,\Vec{r}_2)\,,
\eeq where $\rho^{(1)}(\Vec{r}_1,\Vec{r}_1^\prime)$ and
$\rho^{(2)}(\Vec{r}_1,\Vec{r}_2)$ are the OBMD and the TBD matrices,
respectively, which are  defined as the expectation value of the
operators \beq
\hat{\rho}_1(\Vec{\tilde{r}}_1,\Vec{\tilde{r}}_1^\prime)\,=\,\sum_i\,\delta(\Vec{r}_i-\Vec{\tilde{r}}_1)
        \,\delta(\Vec{r}_i^\prime-\Vec{\tilde{r}}^\prime_1)\,\prod_{j\neq i}
        \,\delta(\Vec{r}_j-\Vec{r}^\prime_j)
\eeq
and
\beq \label{tbd}
\hat{\rho}_2(\Vec{\tilde{r}}_1,\Vec{\tilde{r}}_2)\,=\,\sum_{i<j}\,
        \delta(\Vec{r}_i-\Vec{\tilde{r}}_1)
        \,\delta(\Vec{r}_j-\Vec{\tilde{r}}_2),
        \eeq
\ie \beqy \label{eq22}
\rho^{(1)}(\Vec{r}_1,\Vec{r}_1^\prime)&=&
     \bra{\psi_o}\,\hat{\rho}_1(\Vec{r}_1,\Vec{r}_1^\prime)\,\ket{\psi_o^\prime}\\
\label{eq23}
\rho^{(2)}(\Vec{r}_1,\Vec{r}_2)&=&
     \bra{\psi_o}\,\hat{\rho}_2(\Vec{r}_1,\Vec{r}_2)\,\ket{\psi_o}\,,
\eeqy
where
\beq
\psi_o\,\equiv\,\psi_o(\Vec{x}_1,...,\Vec{x}_A)\,
\eeq
and
\beq
\psi^\prime_o\,\equiv\,\psi_o(\Vec{x}^\prime_1,...,\Vec{x}^\prime_A)\,
\eeq
and a summation over spin and isospin variables  is implicit in Eqs. (\ref{eq22})
and (\ref{eq23}). The knowledge of the OBMD and TBD matrices allows one to calculate,
besides the ground-state energy, other relevant quantities like \textit{e.g.} the
density distribution:
\beq\label{obdrr}
\rho(\Vec{r})\,=\,\rho^{(1)}(\Vec{r}_1=\Vec{r}_1^\prime\equiv\Vec{r})\,,
\eeq
the mean square radius of the distribution:
\beq
\label{rms}
\langle r^2\rangle\,=\,\int d\Vec{r}\,r^2\,\rho(r)
\eeq
and, eventually,  the nucleon momentum distribution , \textit{i.e.} the square of
the WF
in momentum space which, by definition, reads as follows:
\beq
\label{defmomdis}
n(\Vec{k})\,=\,\frac{1}{(2\pi)^3}\,\int d\Vec{r}_1 d\Vec{r}^\prime_1
        \,e^{i\,\Vec{k}\cdot(\Vec{r}_1-\Vec{r}^\prime_1)}\,\rho(\Vec{r}_1,
        \Vec{r}^\prime_1)\,.
\eeq
The normalization of the OBD, OBMD and TBD matrices are as follows
 \beq
\label{nordens}
\int d\Vec{r} \, \rho(\Vec{r})\,=\,A\,,
\eeq
\beq
\label{normnond}
\int d\Vec{r}'_1\,\rho^{(1)}(\Vec{r}_1,\Vec{r}'_1)\,=\,\,\rho^{(1)}(\Vec{r}_1)
\eeq
\beq
\int d\Vec{r}_1\,d\Vec{r}_2\,\rho^{(2)}(\Vec{r}_1,\Vec{r}_2)\,=\,\frac{A(A-1)}{2}\,;
\eeq
with the \textit{sequential relation}
\beq
\int d\Vec{r}_2\,\rho^{(2)}(\Vec{r}_1,\Vec{r}_2)\,=\,\frac{A-1}{2}\,\rho^{(1)}(\Vec{r}_1)
\eeq
linking $\rho^{(2)}(\Vec{r}_1,\Vec{r}_2)$ to $\rho^{(1)}(\Vec{r}_1)$.
Accordingly, the normalization of the nucleon momentum distribution is
\beq
\int d\Vec{k} \, n(\Vec{k})\,=\,A
\eeq

It is useful at this moment to  recall the form of $\rho(\Vec{r}_1)$,
$\rho^{(1)}(\Vec{r}_1,\Vec{r}_1^\prime)$ and $\rho^{(2)}(\Vec{r}_1,\Vec{r}_2)$
predicted by  the SM (or mean field)  approximation. In this case one has
$\psi_o= \phi_o =(A!)^{-1/2}\,
 det \{ \varphi_{\alpha_i}
(\Vec{x}_j)\}$, with the single particle (s.p.) orbitals given by
$\varphi_{\alpha}(\Vec{x}) = \varphi_a(\Vec{r})\,\chi_\sigma^{1/2}\,\xi_\tau^{1/2}$,
where  $\alpha \equiv
 \left\{a;\sigma;\tau\right\} =
\left\{n,l,m;\sigma;\tau\right\}$. For closed shell nuclei one obtains:
\beq
\label{rhoSM}
\rho_{SM}^{(1)}(\Vec{r}_1)\,
           =\,\sum_{\alpha}\left|\varphi_{\alpha}(\Vec{x}_1)\right|^2
           =\,4\,\rho_o(\Vec{r}_1)\,
\eeq
and
\beq
\rho_{SM}^{(1)}(\Vec{r}_1,\Vec{r}^\prime_1)\,
           =\,\sum_{\alpha}\varphi^\star_{\alpha}(\Vec{x}_1)
         \varphi_{\alpha}(\Vec{x}^\prime_1)
         =\,4\,\rho^{(1)}_o(\Vec{r}_1,\Vec{r}^\prime_1)\,
\eeq
where the sum over $\alpha$  runs over the occupied SM states below the Fermi level, and
\beq
\rho_o(\Vec{r}_1)\,=\,\sum_a\left|\varphi_a(\Vec{r}_1)\right|^2
\eeq

\noindent and
\beq
\rho^{(1)}_o(\Vec{r}_1,\Vec{r}^\prime_1)\,=\,\sum_a\varphi^\star_a(\Vec{r}_1)
         \,\varphi_a(\Vec{r}^\prime_1)\,.
\eeq For the  TBD matrix one obtains
\beqy
\rho^{(2)}_{SM}(\Vec{r}_1,\Vec{r}_2)&=&\frac{1}{2}\sum_{\alpha\beta}\Big(
        \,\varphi^\star_\alpha(\Vec{x}_1)\,\varphi^\star_\beta(\Vec{x}_2)
      \,\varphi_\alpha(\Vec{x}_1)\,\varphi_\beta(\Vec{x}_2)\,-\,
        \,\varphi^\star_\alpha(\Vec{x}_1)\,\varphi^\star_\beta(\Vec{x}_2)
      \,\varphi_\beta(\Vec{x}_1)\,\varphi_\alpha(\Vec{x}_2)\Big) \,
      \nn\\
&=&\frac{1}{2}\,4\,\Big(4\,\rho_o(\Vec{r}_1)\,\rho_o(\Vec{r}_2)\,-\,
        \rho^{(1)}_o(\Vec{r}_1,\Vec{r}_2)\,\rho^{(1)}_o(\Vec{r}_2,\Vec{r}_1)\Big)\,,
\eeqy
 where $\rho_o(\Vec{r}_i) =
\rho^{(1)}_o(\Vec{r}_i,\Vec{r}_i)$. When Eq. (\ref{eq22}) is
evaluated with the correlated  wave functions (\ref{psi1}) within
  the $f_6$ approximation,
at  first
order of  the $\eta$-expansion the following expression
 is
obtained:
\beq\label{obrhoeta}
\rho^{(1)}(\Vec{r}_1,\Vec{r}_1^\prime)\,=\,\rho^{(1)}_{SM}(\Vec{r}_1,\Vec{r}_1^\prime)
        \,+\,\rho^{(1)}_{H}(\Vec{r}_1,\Vec{r}_1^\prime)
        \,+\,\rho^{(1)}_{S}(\Vec{r}_1,\Vec{r}_1^\prime)\,,
\eeq
with
\beqy
&&\rho^{(1)}_{H}(\Vec{r}_1,\Vec{r}_1^\prime)  \nonumber\\
       &=& \int\,d\Vec{r}_2\,\Big[H_D(r_{12},r_{1^\prime2})\,\rho^{(1)}_o(\Vec{r}_1,\Vec{r}_1^\prime)
        \,\rho_o(\Vec{r}_2)\,-\,H_E(r_{12},r_{1^\prime2})\,\rho^{(1)}_o(\Vec{r}_1,\Vec{r}_2)
      \,\rho^{(1)}_o(\Vec{r}_2,\Vec{r}_1^\prime)\Big]
      \label{39}
      \eeqy
\beqy
\hspace {-5.0mm} &&\rho^{(1)}_{S}(\Vec{r}_1,\Vec{r}_1^\prime)  \nonumber\\
\hspace {-15.0mm}&&=-\int d\Vec{r}_2d\Vec{r}_3\rho^{(1)}_o(\Vec{r}_1,\Vec{r}_2)\Big[H_D(r_{23})
        \rho^{(1)}_o(\Vec{r}_2,\Vec{r}_1^\prime)\rho_o(\Vec{r}_3)-
H_E(r_{23})
        \rho^{(1)}_o(\Vec{r}_2,\Vec{r}_3)\rho^{(1)}_o(\Vec{r}_3,
        \Vec{r}_1^\prime)\Big]
        \label{40}
\eeqy
where $r_{ij}=|{\bf r}_i -{\bf r}_j|$. The subscripts $H$ and $S$,
 whose meaning will be explained in Section \ref{section7},
  stand for \textit{hole} and \textit{spectator},
respectively,
and
\beq
\label{accadirex}
H^{D(E)}(r_{ij},r_{kl})\,=\,\sum^6_{p,q=1}\,f^{(p)}(r_{ij})\,f^{(q)}(r_{kl})
       \,C^{(p,q)}_{D(E)}(r_{ij},r_{kl})\,-\,C^{(1,1)}_{D(E)}(r_{ij},r_{kl})\,.
\eeq
Here the subscripts $D$ and $E$ stand for \textit{direct}
 and \textit{exchange}, respectively,  and the  coefficients
  $C^{(p,q)}_{D}(r_{ij},r_{kl})$, $C^{(p,q)}_{E}(r_{ij},r_{kl})$, whose explicit expressions
 are  given in Appendix C,
result from the \textit{spin} and \textit{isospin} summation, and  their  explicit
dependence
upon the coordinates originates from the tensor operator.
As for the correlated  TBD matrix, this can be obtained by multiplying
the TBD operator in Eq. (\ref{tbd})
by the operators $\mh^{(n)}_{ij}$ of Eq. (\ref{newuno}); the resulting TBD matrix,
corresponding to the operator  $\mh^{(n)}_{ij}$
 reads as follows:
\beq
\label{pot2}
\rho^{(2)}_{(n)}(\Vec{r}_1,\Vec{r}_2)\,=\,\rho^{(A)}_{(n)}(\Vec{r}_1,\Vec{r}_2)\;+
        \;\rho^{(B)}_{(n)}(\Vec{r}_1,\Vec{r}_2)\;+\;\rho^{(C)}_{(n)}(\Vec{r}_1,\Vec{r}_2)
      \;+\;\rho^{(D)}_{(n)}(\Vec{r}_1,\Vec{r}_2)
      \eeq
\beqy
%\sublabon{equation}
\rho^{(A)}_{(n)}(\Vec{r}_1,\Vec{r}_2)&=&\frac{1}{A(A-1)}\sum^6_{p,q=1}\,
        f^{(p)}_{12}f^{(q)}_{12}\sum^6_{r,s=1}\, \left( ( K^{(r)}_{(p,q)}K^{(s)}_{(r,n)}
        \,A^{(s)}_D-A^{(n)}_D)\,\rho_o(\Vec{r}_1)\,\rho_o(\Vec{r}_2)\,\right.\,\nonumber\\
        &&\left.\ -( K^{(r)}_{(p,q)}K^{(s)}_{(r,n)}\,A^{(s)}_E-A^{(n)}_E)\,
       \rho_o(\Vec{r}_1,\Vec{r}_2)\,\rho_o(\Vec{r}_2,\Vec{r}_1)\right)
             \label{44}
      \eeqy
\beqy \rho^{(B)}_{(n)}(\Vec{r}_1,\Vec{r}_2)&=&\frac{1}{A(A-1)}
        \int d\Vec{r}_3\sum_{\mathcal P}\left(\sum^6_{p,q=1} f^{(p)}_{13}f^{(q)}_{13}
        B^{(p,q)}_{(n),\mathcal P} - B^{1,1}_{(n),\mathcal P}\right)\,\times\nn\\
&&\hspace{2cm}\times\,\rho_o(\Vec{r}_1,\Vec{r}_{\mathcal{P}\{1\}})\,\rho_o(\Vec{r}_2,\Vec{r}_{\mathcal{P}\{2\}})
          \,\rho_o(\Vec{r}_3,\Vec{r}_{\mathcal{P}\{3\}})\,\label{denspot1b}
          \eeqy

\beqy
 \rho^{(C)}_{(n)}(\Vec{r}_1,\Vec{r}_2)&=&\frac{1}{A(A-1)} \int
d\Vec{r}_3
 \sum_{\mathcal P}
        \left(\sum^6_{p,q=1} f^{(p)}_{23}f^{(q)}_{23}
        C^{(p,q)}_{(n),\mathcal P} - C^{1,1}_{(n),\mathcal P} \right)\,\times\nonumber\\
&&\hspace{2cm}\times\,\rho_o(\Vec{r}_1,\Vec{r}_{\mathcal{P}\{1\}})\,\rho_o(\Vec{r}_2,
\Vec{r}_{\mathcal{P}\{2\}})
          \,\rho_o(\Vec{r}_3,\Vec{r}_{\mathcal{P}\{3\}})\,\label{denspot1c}
          \eeqy
\beqy
\rho^{(D)}_{(n)}(\Vec{r}_1,\Vec{r}_2)&=&\frac{1}{A(A-1)}\frac{1}{2}
\int d\Vec{r}_3
        d\Vec{r}_4\,\sum_{\mathcal P}\left(\sum^6_{p,q=1} f^{(p)}_{34}f^{(q)}_{34}
        \sum^6_{r=1} K^{(r)}_{(p,q)} D^{(r)}_{(n),\mathcal P}
         - D^{(1)}_{(n),\mathcal P} \right)\,\times\nonumber\\
&&\hspace{2cm}\times\,\rho_o(\Vec{r}_1,\Vec{r}_{\mathcal{P}\{1\}})\,\rho_o(\Vec{r}_2,\Vec{r}_{\mathcal{P}\{2\}})
       \,\rho_o(\Vec{r}_3,\Vec{r}_{\mathcal{P}\{3\}})\,\rho_o(\Vec{r}_4,\Vec{r}_{\mathcal{P}\{4\}})
         \eeqy

This expression deserves a few explanations; $A^{(n)}_{D(E)}$,
$B^{(p,q)}_{(n),\mathcal P}$, $ C^{(p,q)}_{(n),\mathcal P}$ and
$D^{(r)}_{(n),\mathcal P} $ are the result of the spin-isospin
summations and they are in general function of the coordinates; the
remaining summations over the spatial quantum numbers are then
expressed in terms of combinations of $OBMD$ matrices (see Appendix
C); the subindex ${\mathcal P} $ in these factors stands for all
possible permutations of the states but the \textit{unlinked} one
and the sub-indexes $\mathcal{P}\{i\}$ means the corresponding index
 resulting from the particular permutations; finally, the
matrices $ K^{(r)}_{(p,q)}$ are proper \textit{numerical}
combination of the spin-isospin operators $\mh^{(n)}_{ij}$, and are
defined by the following relation \beq
\mh^{(p)}_{(m,n)}\,\mh^{(q)}_{(m,n)}\,=\,\sum^6_{r=1}\,
K^{(r)}_{(p,q)}\,\mh^{(r)}_{(m,n)} \eeq
Note that even if we are
dealing with two-body correlations and interactions we end up with
three- and four-body operators, due to the fact that, \textit{e.g.}
terms like $\mh_{(1,2)}\mh_{(1,3)}$, cannot be further reduced.
Thus,
 the first order $\eta$-expansion for the energy gets contributions
from up to four-body clusters.

>From the definition of the nucleon momentum distribution, \ie{ }
Eq.(\ref{defmomdis}), we can obtain the expectation value of the
kinetic energy operator as follows:
\beq \label{kin1}
\langle\hat{T}\rangle\,=\,\frac{\hbar^2}{2\,M_N}\,\int\,d\Vec{k}\,k^2\,n(\Vec{k})\,.
\eeq
 and Eq. (\ref{hmatrix}), finally becomes \beq\label{newhmat}
E_o\,=\,\,\frac{\hbar^2}{2\,M_N}\,\int\,d\Vec{k}\,k^2\,n(\Vec{k})\,+
        \,\sum_n\;\int\;d\Vec{r}_1 d
        \Vec{r}_2\;v^{(n)}(r_{12})\rho^{(2)}_{(n)}(\Vec{r}_1,\Vec{r}_2)\,,
\eeq
with $\rho^{(2)}_{(n)}(\Vec{r}_1,\Vec{r}_2)$ given by Eq. (\ref{pot2}).
This is the final expression which has been used to calculate the ground-state energy
 by the following procedure: we have calculated at the same order both $n(k)$ and
  $\rho^{(2)}_n$, then by placing them in Eq. (\ref{newhmat}) and performing the summation
  over $n$ the ground state energy $E_0$  is obtained. Calculations have been performed
  with   a given, fixed form for the correlation functions,
  and considering as variational parameters, the parameters of the mean
  field wave functions. To begin with, in the next Section the results
  of a benchmark calculation aimed at investigating the convergence of the expansion
  will be presented.

\subsection{A benchmark calculation for $^{16}O$ ans $^{40}Ca$: comparison
  between the $\eta$-expansion and the Fermion-Hyper-Netted-Chain /
  Single Operator Chain (FHNC/SOC) approach with truncated $V8^\prime$
  and $U14$ interactions}

\begin{figure}[!hp]
  \centerline{
    \epsfysize=0.5\textwidth{\epsfbox{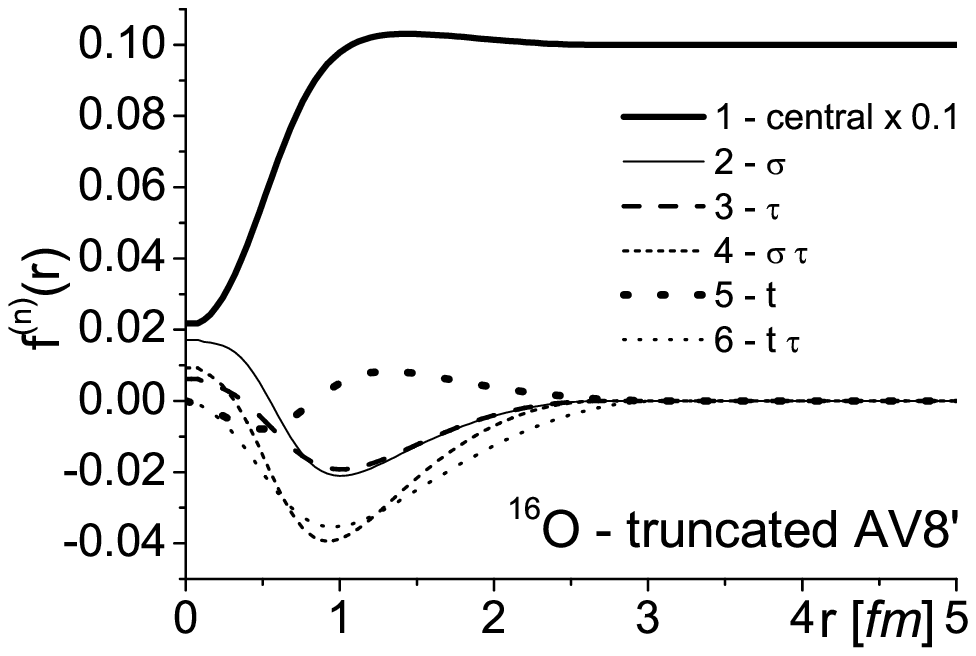}}}
  \caption{The correlation functions for $^{16}O$ corresponding to the
    truncated Argonne $AV8^\prime$ interaction \cite{pud01} used in the benchmark
    calculation (After Ref. \cite{30a}).}
  \label{Fig1}
\end{figure}

\begin{figure}[!hcp]
  \centerline{
    \epsfysize=0.5\textwidth{\epsfbox{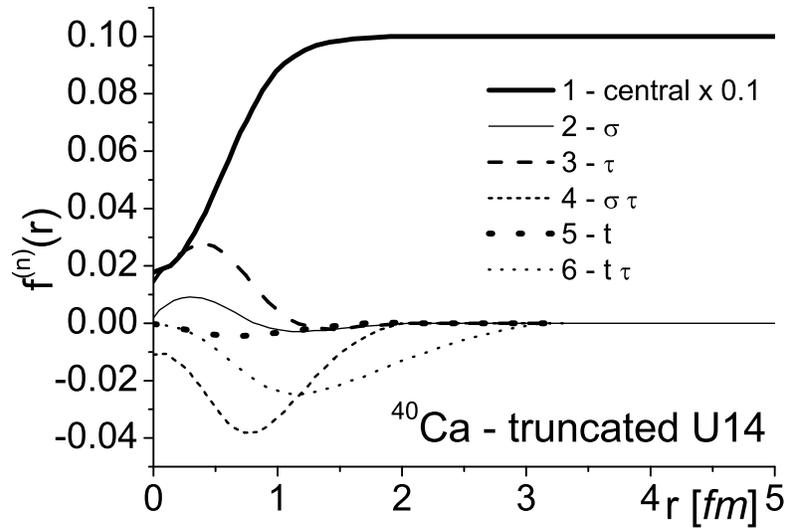}}}
  \caption{The  correlation functions for $^{40}Ca$ corresponding to the
    truncated Urbana $U14$ interaction \cite{u14} used in the benchmark calculation
    \cite{pud01} (After Ref. \cite{30a}).}
  \label{Fig2}
\end{figure}

In order to investigate the convergence of the $\eta$-expansion, we
have performed a benchmark calculation consisting in a comparison of
Eq. (\ref{newhmat}) with the energy predicted by the FHNC/SOC
approach. Namely,  we have calculated the ground-state properties of
$^{16}O$ and $^{40}Ca$ using the first six components of the
$V8^\prime$ \cite{pud01} and $U14$ \cite{30b} interactions,
respectively (these model interactions are usually referred to as
the {\it truncated} $V8^\prime$ and $U14$ interactions). The results
we have obtained are compared with the results obtained with
FHNC/SOC using the same interaction, the same  mean field WF's, and
the same correlation functions. The six correlation functions used
in the calculation for $^{16}O$ , corresponding to the $V8^\prime$
interaction,  are shown in Fig. \ref{Fig1}, and the results of the
energy calculation  are presented in Tables \ref{Table1} and
\ref{Table2}. It can be seen that the cluster expansion results are
very similar to the ones provided by the FHNC/SOC method;
particularly worth being mentioned is the almost identical value of
the mean kinetic energy, which means that the nucleon momentum
distributions predicted by the two methods are very similar. The
results of the calculation for $^{40}Ca$, corresponding to the
truncated $U14$ interaction \cite{30b} and to the mean field and
correlation functions shown in Fig. \ref{Fig2}, are presented in
Table \ref{Table3} where they are compared with the results of the
FHNC/SOC approach of Ref. \cite{fab02}. Being the mean field wave
functions and correlation functions the same in the two
calculations, any difference between our results and those of Ref.
\cite{fab02}has to be ascribed, as in the case of $^{16}O$, to the
contributions that are left out in the cluster expansion. It can be
seen that the difference between the two approaches is larger in
$^{40}Ca$ than in the $^{16}O$ case, the largest difference arising
from the spin-isospin interaction, which as a matter of fact is of
longer range in $^{40}Ca$ (cf. Fig. \ref{Fig2}).

To sum up, it seems that the convergence of the $\eta$-expansion for the ground-state
energy  is
a satisfactory one.

\section{Application of the $\eta$-expansion to the Nuclei $^{16}O$ and $^{40}Ca$: the ground-state energy, radius
  and densities with the full $V8^\prime$ interaction}

\begin{figure}[!hp]
  \centerline{
    \epsfysize=0.5\textwidth\epsfbox{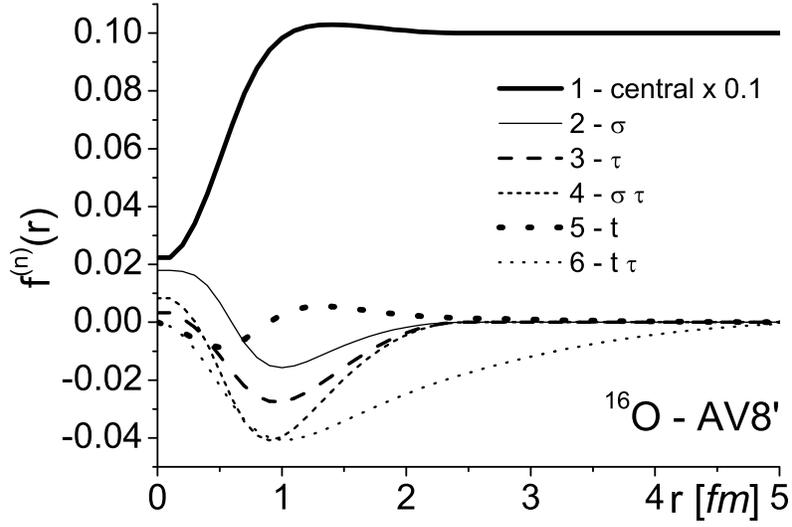}}
  \caption{The correlation functions for $^{16}O$ corresponding to the
  $AV8^\prime$ interaction (After Ref. \cite{fab01}).}
  \label{Fig3}
\end{figure}

\begin{figure}[!hp]
  \centerline{
    \epsfysize=0.5\textwidth\epsfbox{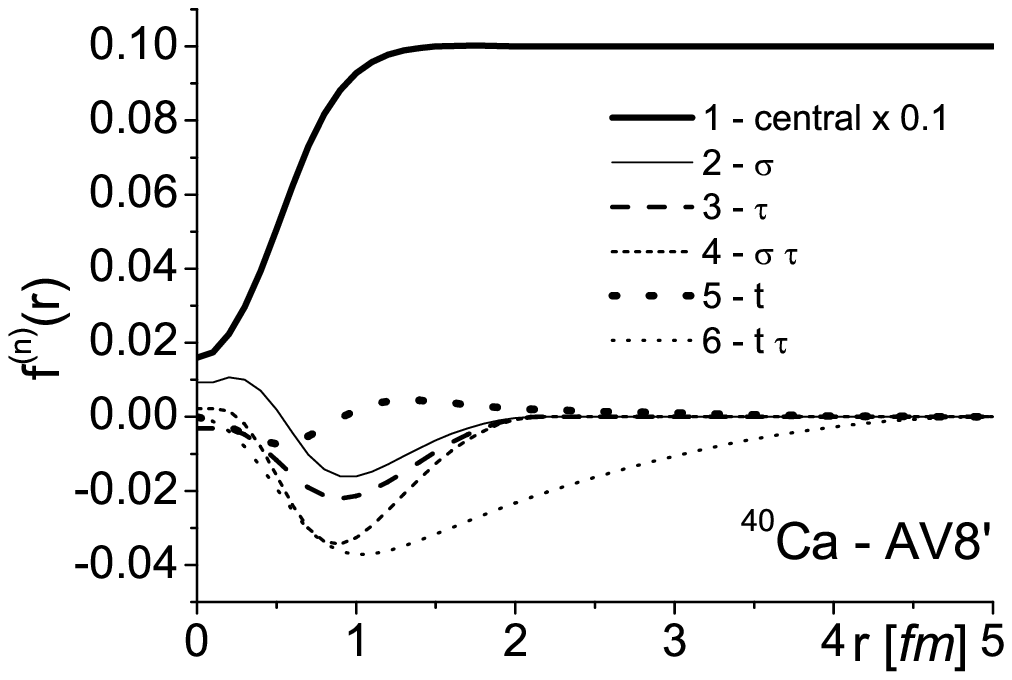}}
  \caption{The correlation functions for $^{40}Ca$ from \cite{fab01},
    corresponding to the  $AV8^\prime$
    interaction \cite{pud01} (After Ref. \cite{fab01}).}
  \label{Fig4}
\end{figure}

In Ref. \cite{fab01},  using the full $V8^\prime$
interaction which includes the spin-orbit contributions $v_7$ and $v_8$,
 several  ground-state properties
of $^{16}O$ and $^{40}Ca$ have been calculated within the FHNC/SOC approach, namely
the ground state energy and
the density and momentum distributions.
For this reason, we have also calculated the ground-state properties of $^{16}O$
and $^{40}Ca$ using the $\eta$-expansion and the correlation functions of
 Ref. \cite{fab01}
which are shown in Fig. \ref{Fig3} and \ref{Fig4}.
The FHNC/SOC calculation of  Ref. \cite{fab01}   was performed within the
$f_6$ approximation. We have also used  such an
approximation but, unlike Ref. \cite{fab01}, we have
 disregarded the $v_7$ and $v_8$ components of the
$V8^\prime$ interaction. For such a reason a direct comparison of
the results for the potential energy is not possible, whereas
a comparison of the average kinetic energy is fully meaningful.
The results
of the comparison are presented in Tables \ref{Table5} and \ref{Table6}
for $^{16}O$,  and \ref{Table7} and \ref{Table8} for $^{40}Ca$. The most striking
feature of the correlation functions obtained in Ref. \cite{fab01} is
 the long tail
of the tensor-isospin correlation function $f^{6}$, which is expected to affect
the convergence of the cluster expansion. As a matter of fact,
 it can be seen that
whereas the difference in $\langle T\rangle$ are of the same order
as in the benchmark calculation, for  $\langle V\rangle$ the
situation is not as good as for $\langle T\rangle$. The reason
should probably be ascribed to the  long tail in $f^{6}$, and the
spin-orbit term in the interaction which is dropped in the present
calculation. As far as the latter is concerned, we have estimated
the effect of the inclusion of the angular momentum dependent terms,
by using the nuclear matter results of Ref.\cite{fab01}, and the
discrepancy for the nuclear matter case seems to be consistent with
the discrepancy we found.

The results we have obtained deserve the following comments

\begin{itemize}
\item[i)]{ We have compared our results obtained with the truncated
        $V8^\prime$ and the $f_6$ approximation but using the full $V8^\prime$
      potential which includes the $v_7$ and $v_8$ components. Since in
      both calculations the same mean field WF and correlations functions
      have been used, the differences between the two results have to be
      ascribed to the terms left out in the cluster expansion. Our
      estimate of the contribution of the $v_7$ and $v_8$, based on nuclear
      matter results, shows that this seems indeed to be the case;}
\item[ii)]{the average  kinetic energy obtained
        in  \cite{fab01} agrees with the one obtained by our approach;
 we will indeed show that the momentum distribution, from which the
      kinetic energy is obtained, (see Eq. (\ref{kin1})), is in very
      good agreement with the one obtained in \cite{fab01}. Some discrepancies are
      still present as far as the potential energy is concerned, but obtaining a
      full
      agreement  between the lowest order cluster expansion and the FHNC/SOC
      approaches is illusory.}
\item[iii)]{the overall value of the ground-state energy obtained in this
        section is reasonably closer to the experimental one ($\simeq 8$ $MeV$
        per nucleon) and it appears that the $\eta$-expansion provides a reasonable
      WF as far as the expectation value of the Hamiltonian is concerned.}
\end{itemize}

\normalsize

%%%%%%%%%%%%%%%%%%%%%%%%%%%%%%%%%%%%%%%%%%%%%%%%%%%%%%%%%%%%%%%%%%%%%%%%%%%%%%%%%%%%%

\begin{figure}[!hp]
  \centerline{
    \epsfysize=0.45\textwidth\epsfbox{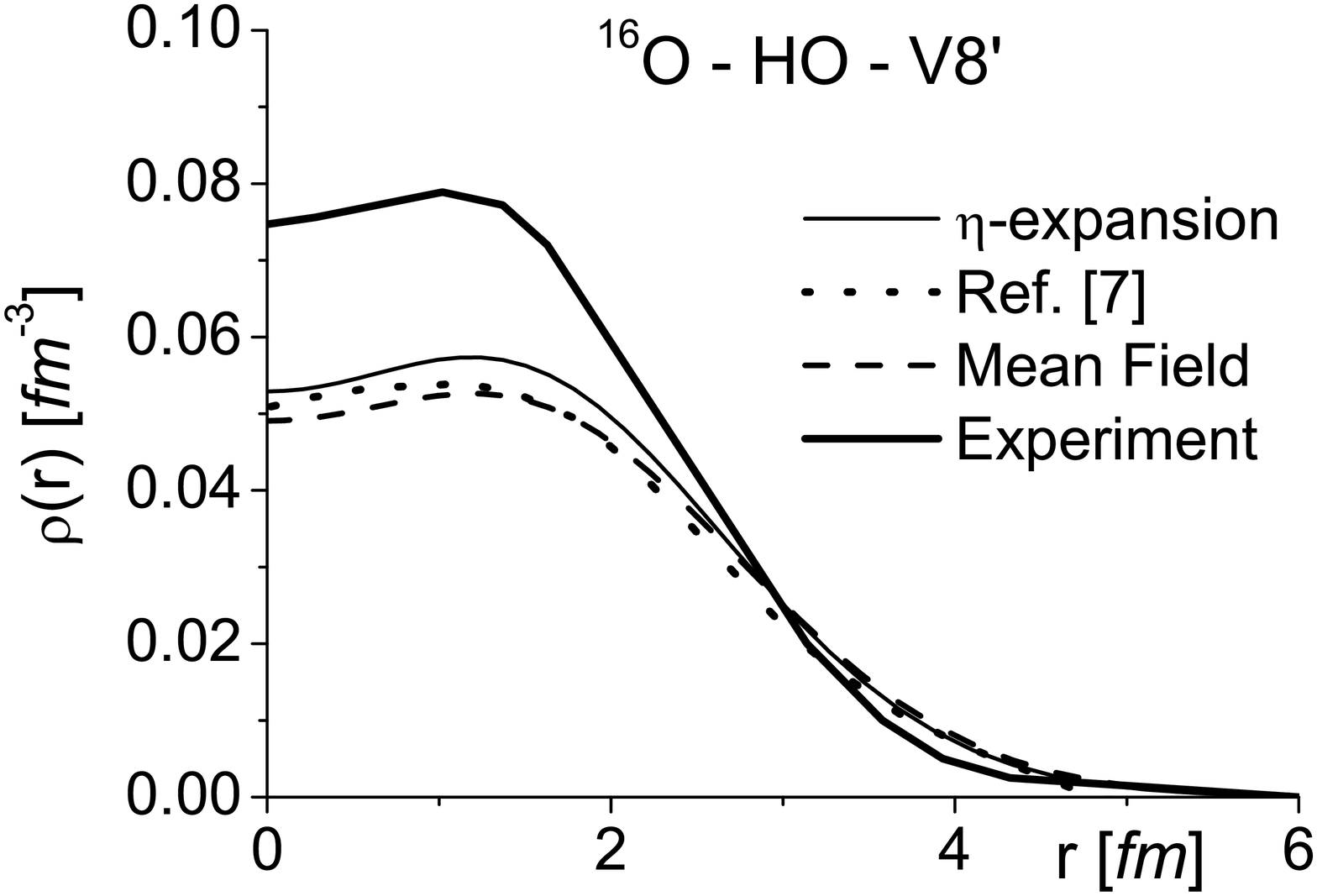}}
  \centerline{
    \epsfysize=0.45\textwidth\epsfbox{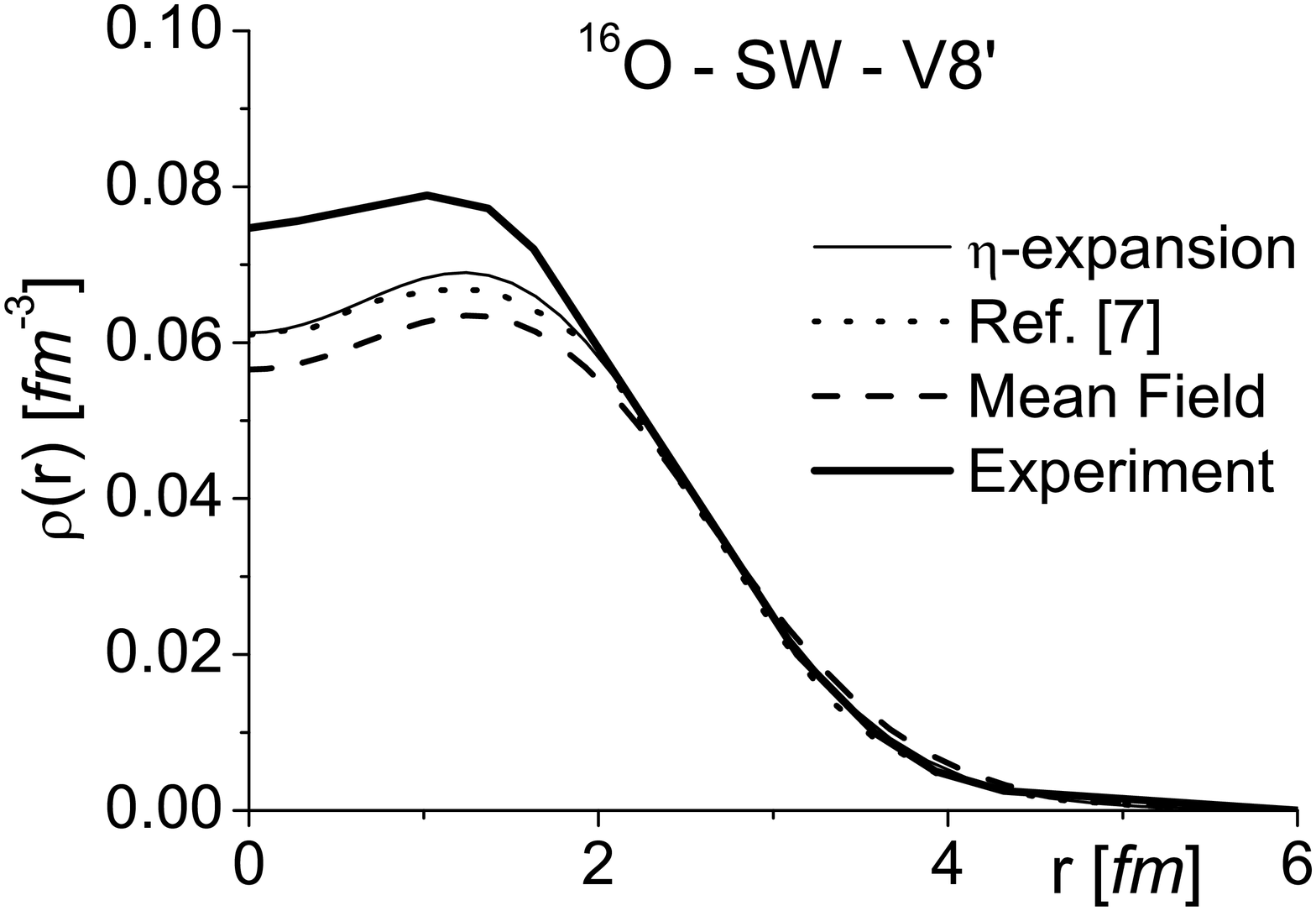}}
  \caption{The charge density of $^{16}O$. \textit{Thick full}:
      experimental density. \textit{Thin full}: results of the
      $\eta$-expansion with harmonic oscillator (\textit{HO,
      top}) and Saxon-Woods (\textit{SW, bottom})
       wave functions and
      correlations functions shown in Fig. \ref{Fig3}. The wave function parameters
      correspond to the minimization  of the ground-state energy.
      \textit{Dots}: mean field density obtained by setting $f^{(1)} = 1$,
       $f^{(n\neq 1)} = 0$.
      The charge density is  obtained by folding the matter
      density with the charge density of the proton
      and correcting for the center-of-mass motion effects. The value of
       the rms radius is ${\langle r^2\rangle}^{1/2} = 3.07$ $fm$,
      with HO wave functions,  and ${\langle r^2\rangle}^{1/2} = 2.85$ $fm$, with SW
      wave functions.
      The value of the HO  parameter is $a = 2.00$ $fm$ and the parameters
      of the SW well are  $V_o = 42.0$ $fm$, $R_o = 3.6$
      $fm$ and $a_o = 0.55$ $fm$.
      The density normalization is $\int d\Vec{r} \rho(r) = Z$.}
  \label{Fig5}
\end{figure}

\begin{figure}[!hp]
  \centerline{
    \epsfysize=0.6\textwidth\epsfbox{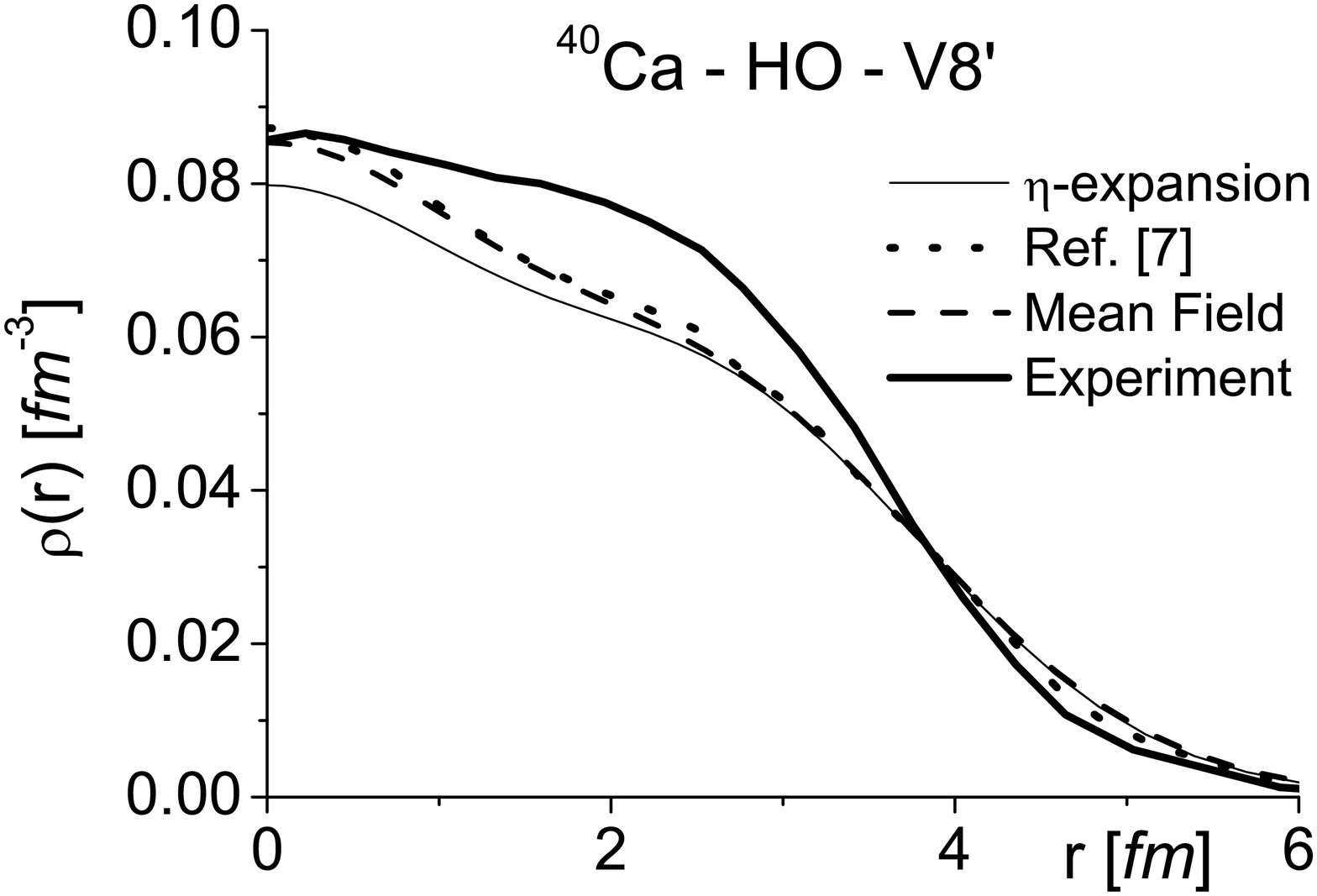}}
  \centerline{\hspace{3mm}
    \epsfysize=0.6\textwidth\epsfbox{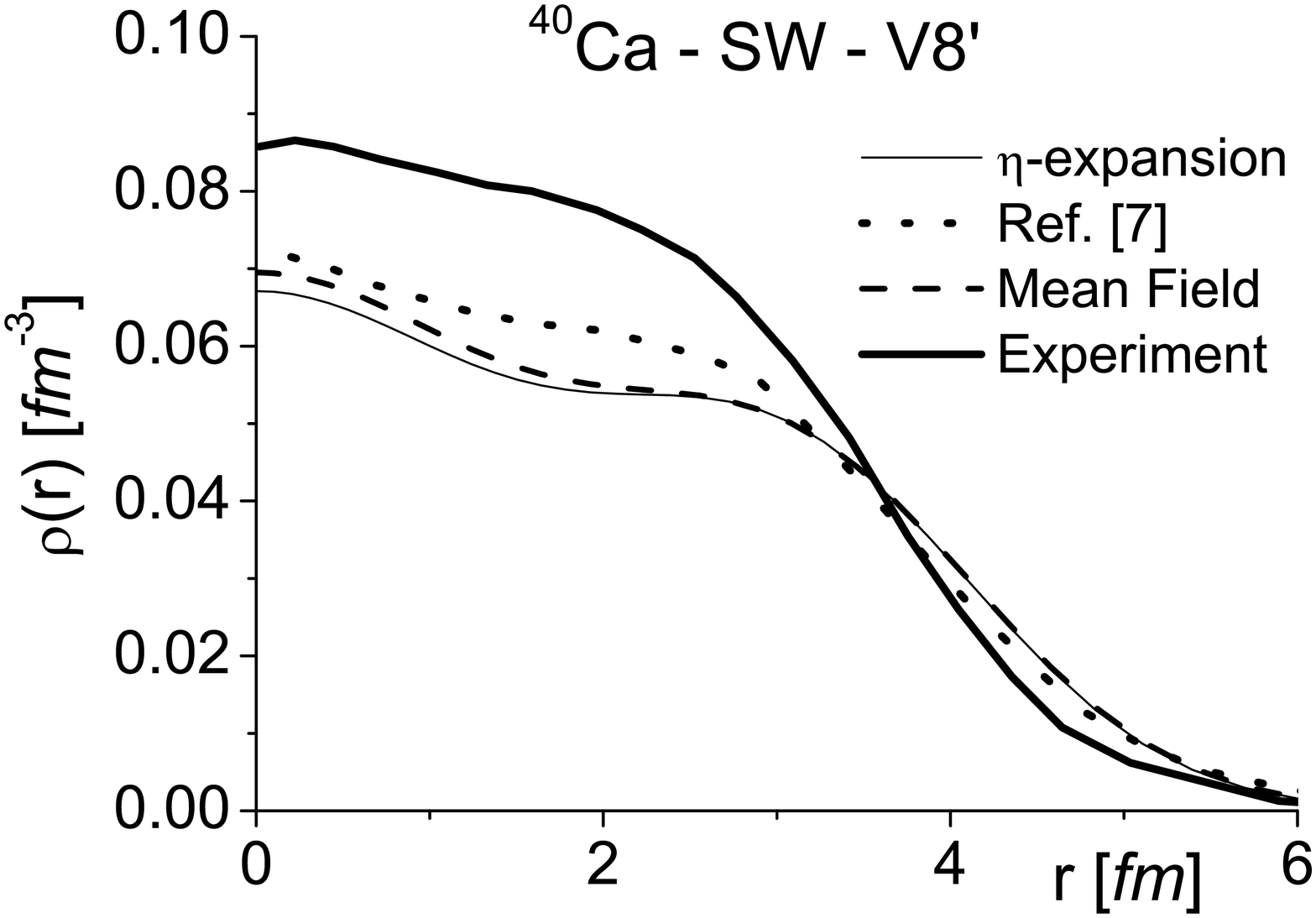}}
  \caption{The same as in Fig. \ref{Fig5}, but for $^{40}Ca$, and correlation
    functions from Fig. \ref{Fig4}. The value of
    the rms radius is ${\langle r^2\rangle}^{1/2} = 3.72$ $fm$,
    with HO wave functions, and ${\langle r^2\rangle}^{1/2} = 3.75$ $fm$, with SW
    wave functions;
    the value of the HO  parameter is $a = 2.10$ $fm$, and the parameters
    of the SW well are $V_o = 50.0$ $fm$, $R_o = 5.3$
    $fm$ and $a_o = 0.53$ $fm$.}
  \label{Fig6}
\end{figure}

By letting $\Vec{r}_1=\Vec{r}_1^\prime\equiv\Vec{r}$ in Eq.
(\ref{obrhoeta}), the matter density at first order of the
$\eta$-expansion, is obtained, \ie \beqy
\rho(\Vec{r})&=&4\,\rho_o(\Vec{r})\,
        \nn\\
&&\!\!\!\!\!\!\!\!\!\!\!\!\!\!\!\!\!\!\!
+\int\,d\Vec{r}_2\,\Big[H_D(r_{12})\,\rho_o(\Vec{r})\,\rho_o(\Vec{r}_2)
        \,-\,H_E(r_{12})\,\rho^{(1)}_o(\Vec{r},\Vec{r}_2)\,\rho^{(1)}_o(\Vec{r}_2,\Vec{r})\Big]\,\nn\\
&&\!\!\!\!\!\!\!\!\!\!\!\!\!\!\!\!\!\!\!\!\!\!\!
        -\int\,d\Vec{r}_2d\Vec{r}_3\,\rho^{(1)}_o(\Vec{r},\Vec{r}_2)\Big[H_D(r_{23})
        \,\rho_o(\Vec{r}_2,\Vec{r})\,\rho_o(\Vec{r}_3)\,-\,H_E(r_{23})
        \,\rho^{(1)}_o(\Vec{r}_2,\Vec{r}_3)\,\rho^{(1)}_o(\Vec{r}_3,\Vec{r})\Big]\,.\label{rhodens}
\eeqy
The charge density is obtained by convoluting $\rho(\Vec{r})$
with the charge density of the proton and by correcting for the center-of-mass
motion (see \textit{e.g.}{ }\cite{30c}). Using the mean field WF and the correlation
functions obtained from the ground-state energy calculation with the full $V8'$
 interaction (cf. Figs. 3 and 4   and Tables V-VIII), the densities shown in Figs. \ref{Fig5}
  and \ref{Fig6} for $^{16}O$ and for $^{40}Ca$,
respectively, are obtained.
The results presented in Fig. \ref{Fig5} and Fig. \ref{Fig6} clearly show that
the charge density calculated within the first order $\eta$-expansion
agrees very well with the results obtained in \cite{fab01} within the
FHNC/SOC approach, which indicates a very good convergence of the
$\eta$-expansion as far as the density is concerned.

\begin{figure}[!hp]
  \centerline{
    \epsfysize=0.6\textwidth\epsfbox{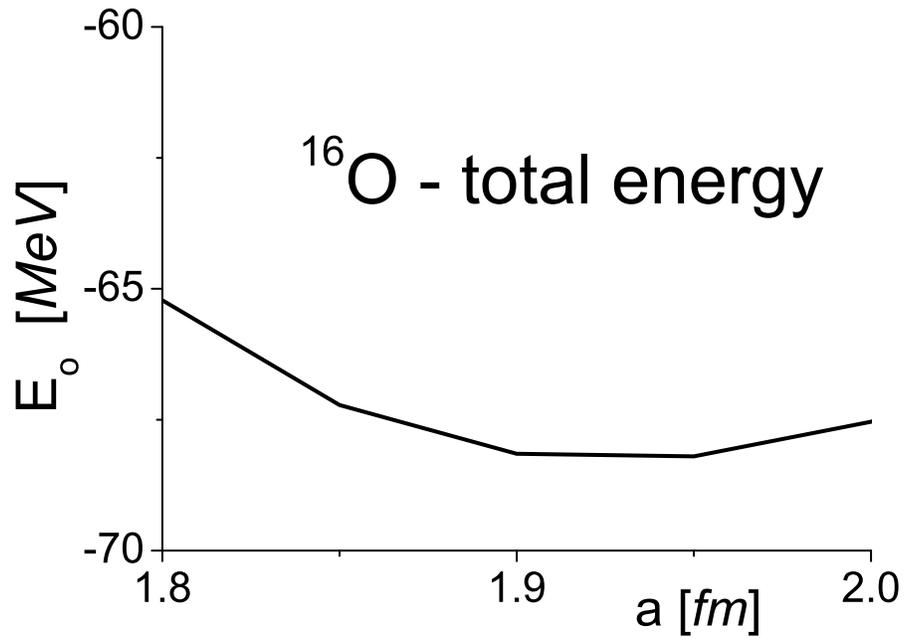}}
  \caption{The ground-state energy of $^{16}O$
      versus the harmonic oscillator parameter $a$
      calculated with the $\eta$-expansion  and the $V8'$ interaction,
      using the correlation functions shown in Fig.3.}
  \label{Fig7}
\end{figure}

\begin{figure}[!hp]
  \centerline{
    \epsfysize=0.6\textwidth\epsfbox{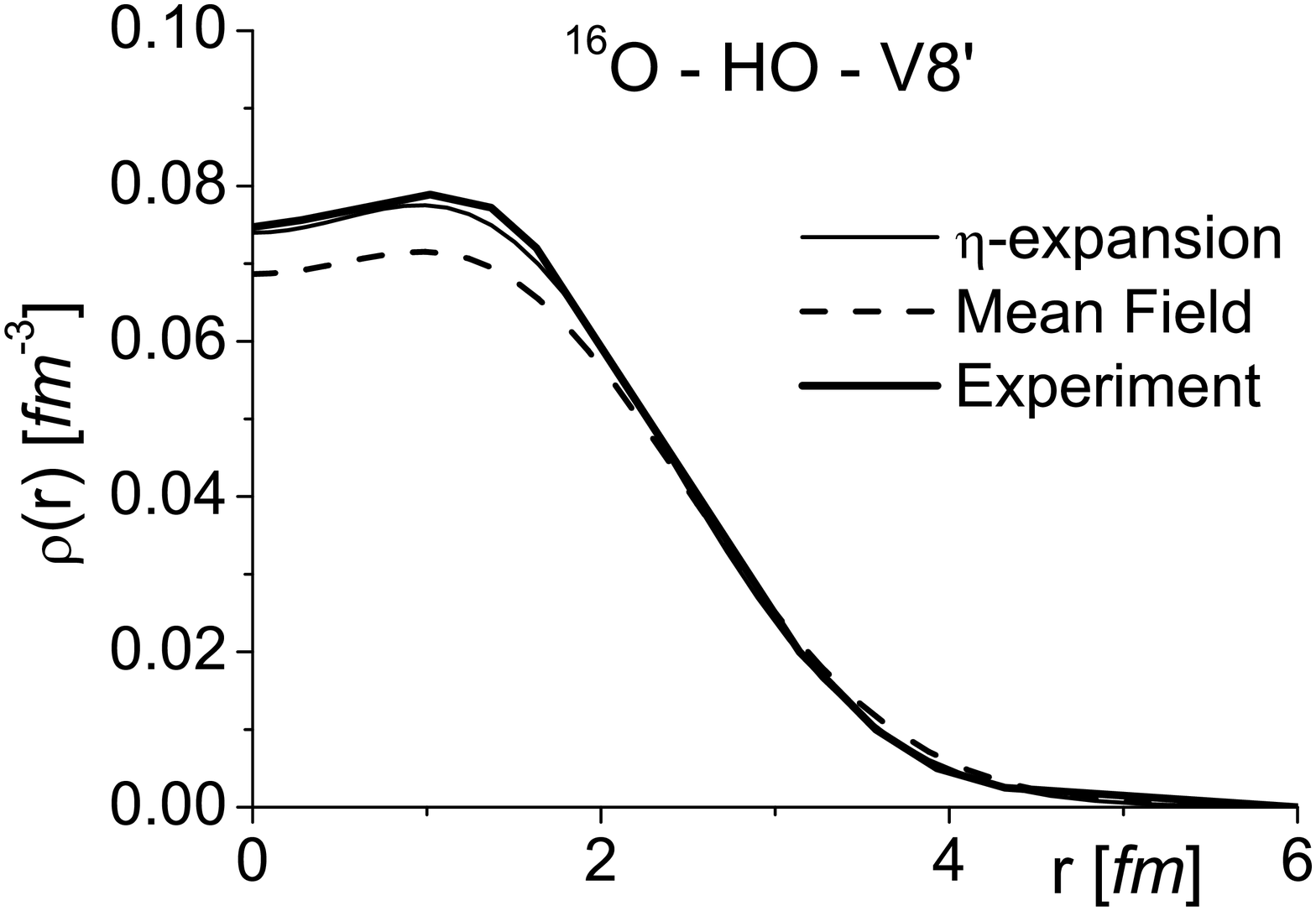}}
  \centerline{
    \epsfysize=0.6\textwidth\epsfbox{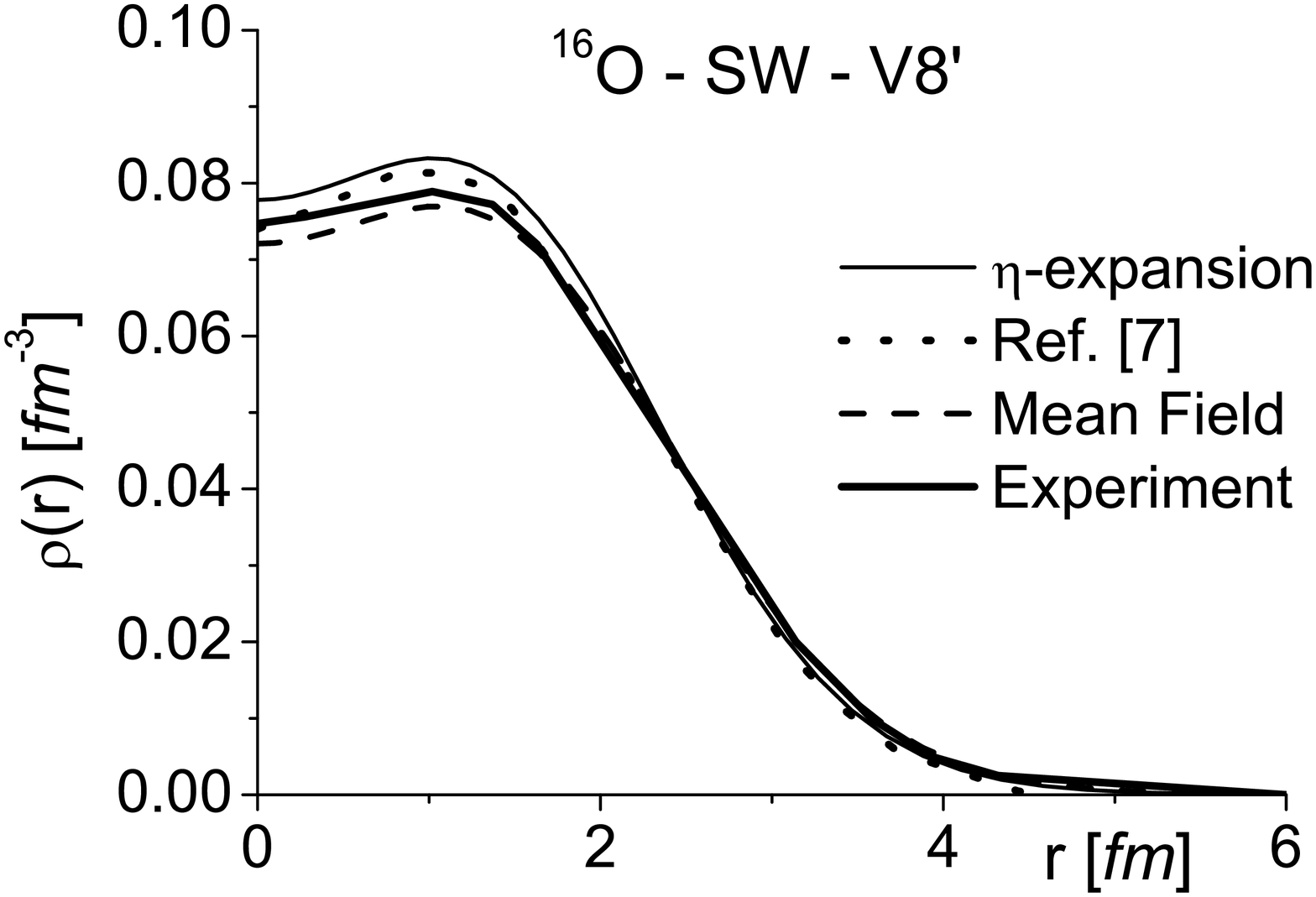}}
  \caption{The same as in Fig, \ref{Fig5}, but with mean field wave functions
   chosen so as to
    better reproduce the experimental density.
    The value of the rms  radius is
    ${\langle r^2\rangle}^{1/2} = 2.73$ $fm$,  with HO wave functions, and
    ${\langle r^2\rangle}^{1/2} = 2.71$ $fm$,  with SW wave functions. The value
    of the  HO  parameter is $a = 1.81$ $fm$, and the
    parameters of the SW well are  $V_o = 53.0$ $fm$,
    $R_o = 3.45$ $fm$ and $a_o = 0.7$ $fm$.}
  \label{Fig8}
\end{figure}

\begin{figure}[!hp]
  \centerline{
    \epsfysize=0.6\textwidth\epsfbox{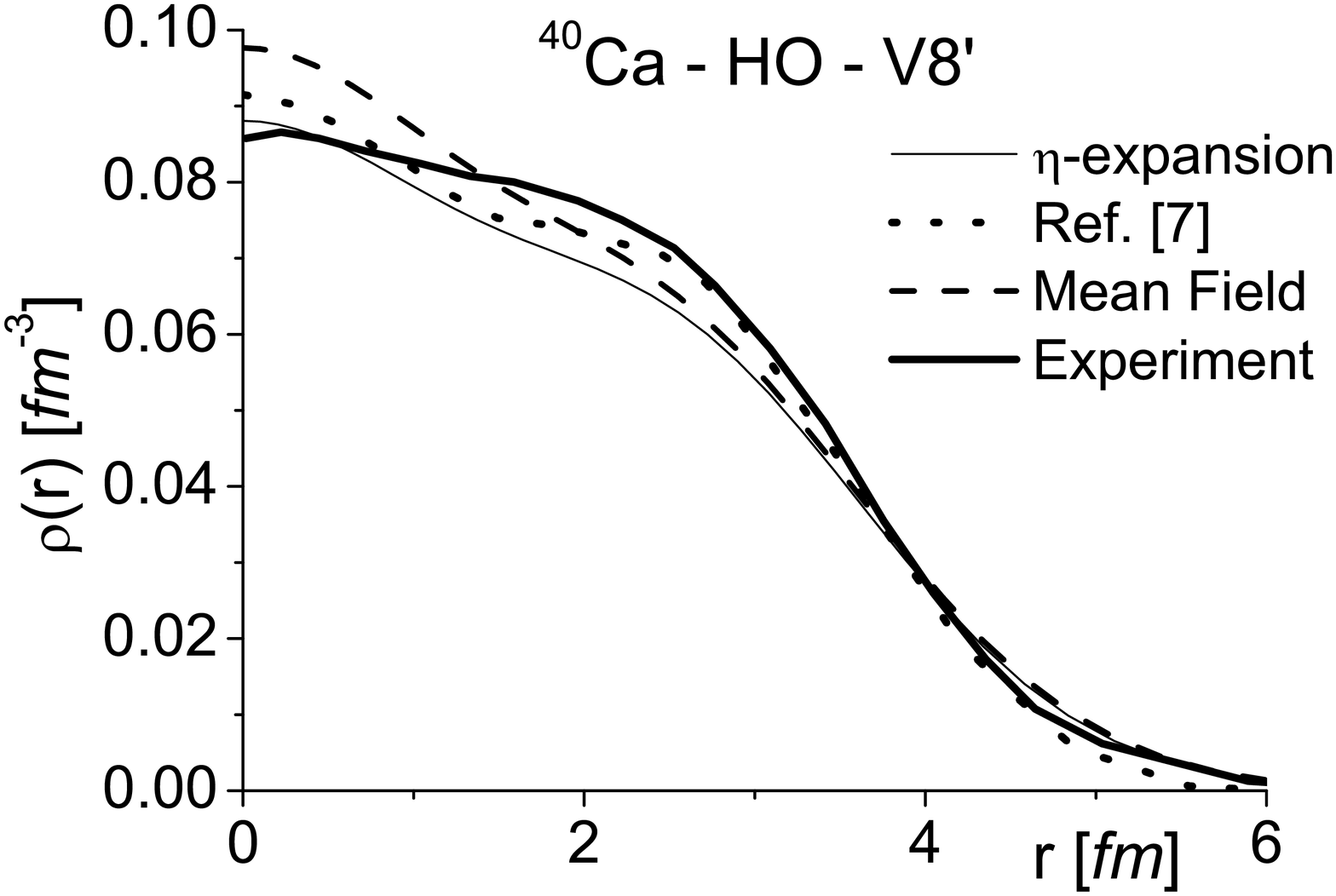}}
  \centerline{\hspace{6mm}
    \epsfysize=0.6\textwidth\epsfbox{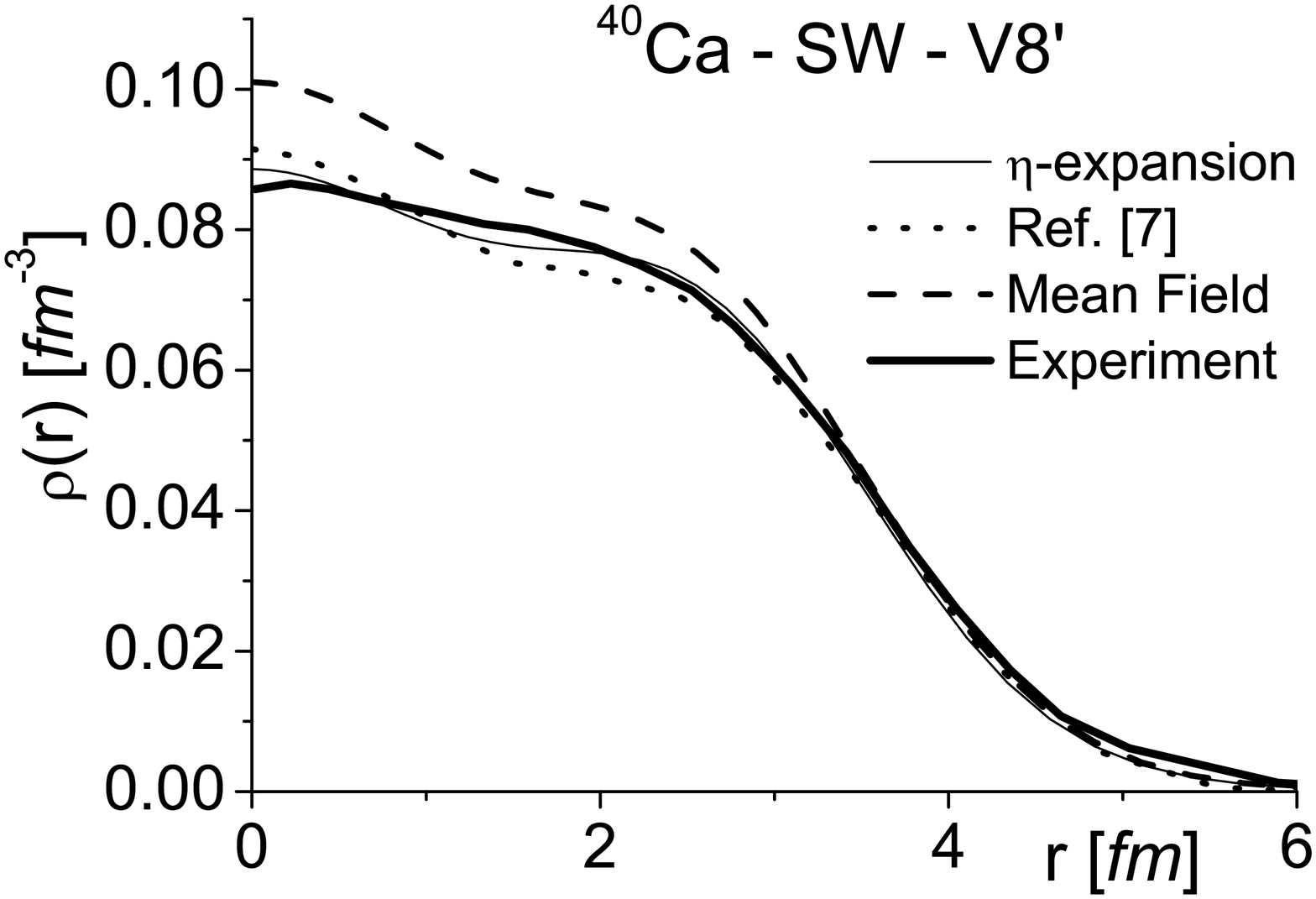}}
  \caption{The same as in Fig. \ref{Fig6}, but for $^{40}Ca$.
    The value of the rms  radius is
    ${\langle r^2\rangle}^{1/2} = 3.56$ $fm$,  with HO wave functions, and
    ${\langle r^2\rangle}^{1/2} = 3.34$ $fm$,  with SW wave functions. The value
    of the HO  parameter is $a = 2.00$ $fm$ and the parameters
    of the SW well are  $V_o = 50.0$ $fm$, $R_o = 5.$ $fm$
    and $a_o = 0.515$ $fm$.}
  \label{Fig9}
\end{figure}

It should however be pointed out, that,  as first found in \cite{fab01},
the density calculated with mean field WF
which minimizes the ground-state energy, strongly disagree with the
experimental density. To cure such a problem, following Ref. \cite{fab01}, we  have
recalculated the density varying the mean field parameters to obtain an agreement with
the experimental density. We take advantage of the fact that, as shown in Fig.
\ref{Fig7}, the energy minimum calculated within the $\eta$-expansion is a rather
 shallow one. The results are shown in Figs. \ref{Fig8} and \ref{Fig9}, and the comparison with
the results of Ref. \cite{fab01} demonstrate
once again the good convergence of the $\eta$-expansion.

\begin{figure}[!hp]
  \centerline{
    \epsfysize=0.6\textwidth\epsfbox{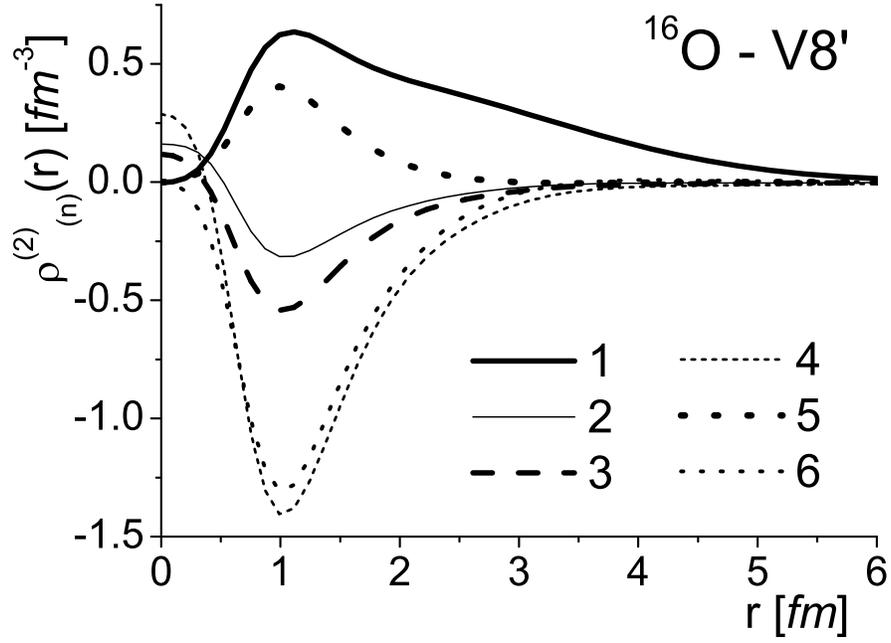}}
  \centerline{
    \epsfysize=0.6\textwidth\epsfbox{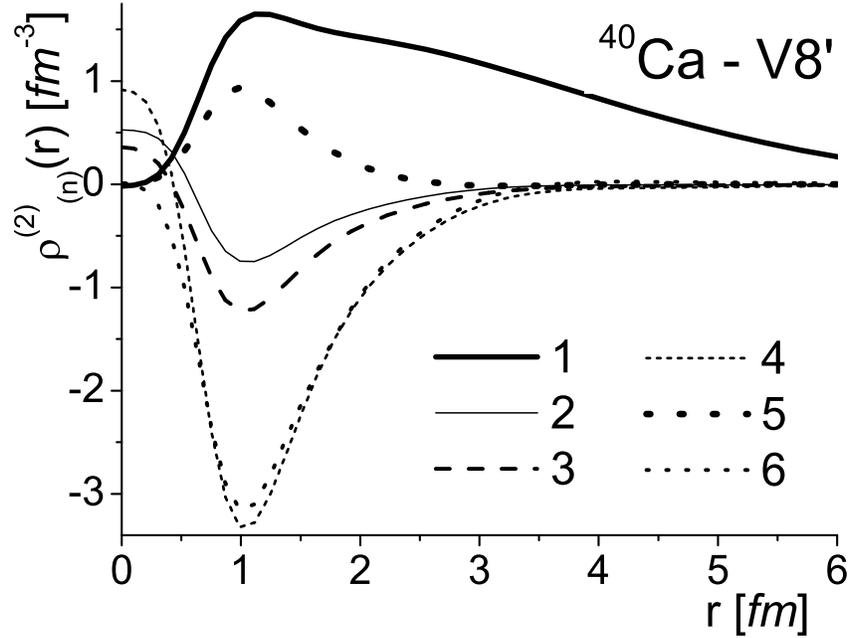}}
  \caption{The two-body densities (Eq. (\ref{pot2})) of $^{16}O$ and $^{40}Ca$
    corresponding to the correlation functions of Fig. \ref{Fig3} ($^{16}O$)
    and Fig. \ref{Fig4} ($^{40}Ca$); the quantities in the figure are integrated
    over the center of mass coordinate (see Eq. (\ref{rho2cm})).
    The splitting of the two-body density in $n$ different quantities is explained
    in Eq. (\ref{pot2}) and the following text.}
  \label{Fig10}
\end{figure}

The six different two-body densities distributions (Eqs.
(42)-(46)) corresponding to the first six correlation operators,
are shown in Fig. \ref{Fig10} for $^{16}O$ (\textit{top}) and
$^{40}Ca$ (\textit{bottom}): each of these densities couples with
the corresponding component of the realistic potential to give the
potential energy expectation value.
 Note that the quantities shown in
Fig. \ref{Fig10} are integrated over the center of mass variable , i.e.

\beq
\label{rho2cm}
\rho^{(2)}_{(n)}(r)\,=\,4 \pi\,\int\,d\Vec{R}\,\rho^{(2)}_{(n)}\left
(\Vec{R}=\frac{1}{2}(\Vec{r}_1+\Vec{r}_2),
  r\equiv| \Vec{r}| = |\Vec{r}_1-\Vec{r}_2|\right)\,.
\eeq

%%%%%%%%%%%%%%%%%%%%%%%%%%%%%%%%%%%%%%%%%%%%%%%%%%%%%%%%%%%%%%%%%%%%%%%%%%%%%%%%%%%%%%%%%%

\section{Application of the $\eta$-expansion to the nuclei $^{16}O$ and $^{40}Ca$: the nucleon momentum distributions}

\begin{figure}[!hp]
  \centerline{
    \epsfysize=0.45\textwidth\epsfbox{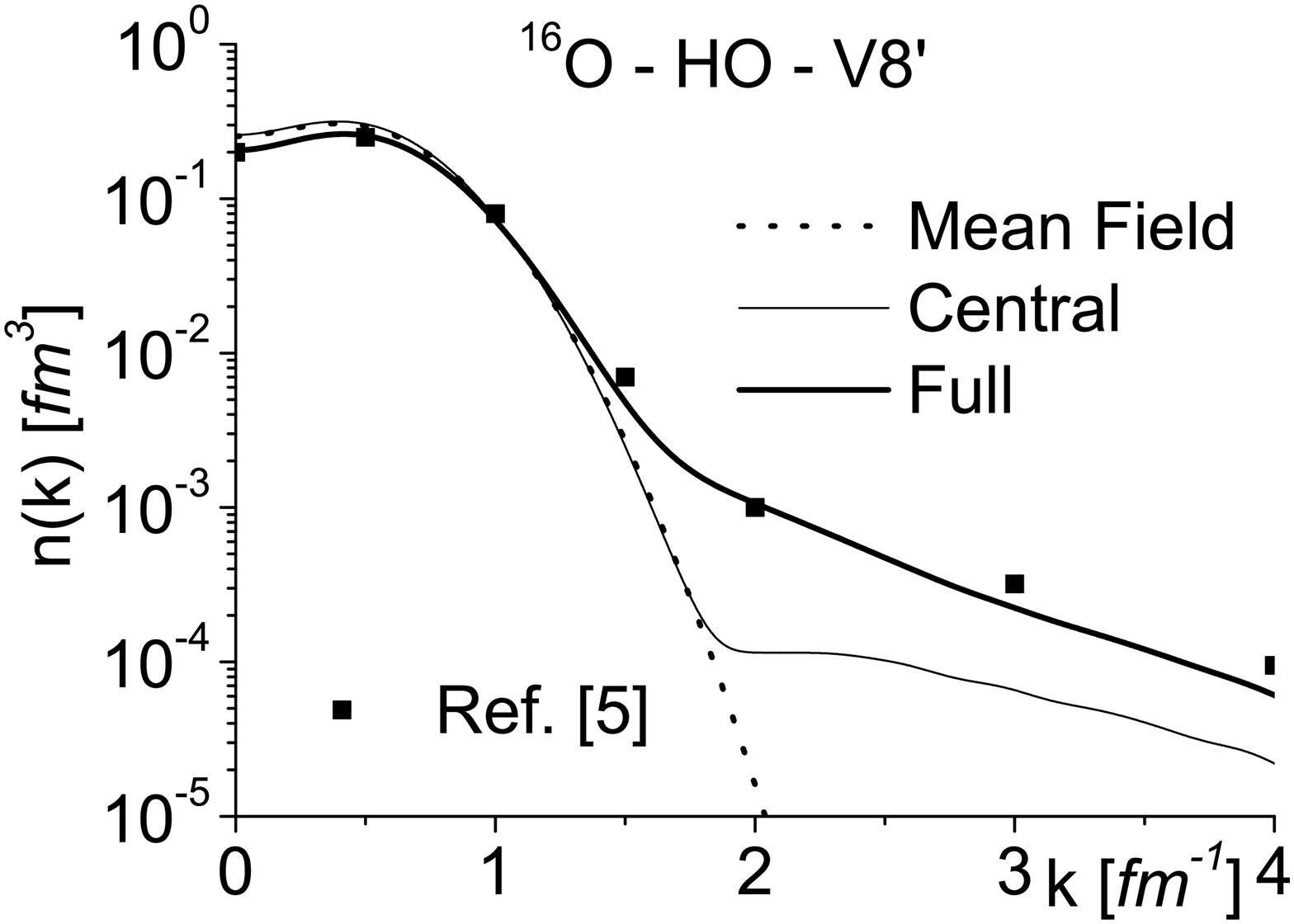}}
  \centerline{
    \epsfysize=0.45\textwidth\epsfbox{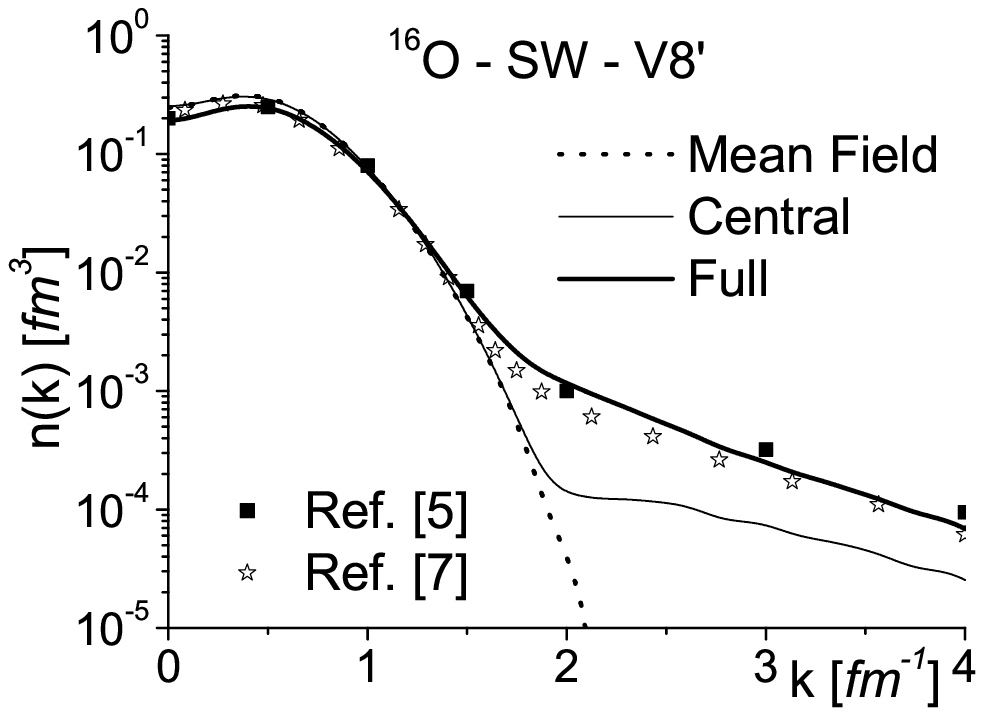}}
  \caption{The momentum distributions of $^{16}O$ corresponding to harmonic
    oscillator (\textit{top}) and Saxon-Woods (\textit{bottom}) wave functions giving
    the best density shown in Fig. \ref{Fig8}.
    The full thin line includes only the central correlation function, whereas the
    thick full line includes all of them. Our results
    are compared with the results of Ref. \cite{fab01} (\textit{stars}), obtained
    with the same correlation functions. The results of Ref. \cite{pie01} obtained
    within the VMC approach  using
    the AV14  interaction are also shown by full squares.
    The value of the kinetic energy obtained  by integrating $n(k)$ are:
    $\langle T\rangle=$ 297.87
    (\textit{central}, HO), $\langle T\rangle=$ 476.55 $MeV$ (\textit{full}, HO);
    $\langle T\rangle=$ 306.99 (\textit{central}, SW) and $\langle T\rangle=$ 494.48
    $MeV$ (\textit{full}, WS). The normalization of $n(k)$ is $\int d\Vec{k}n(k)=1$.}
  \label{Fig11}
\end{figure}

\begin{figure}[!hp]
  \centerline{
    \epsfysize=0.6\textwidth\epsfbox{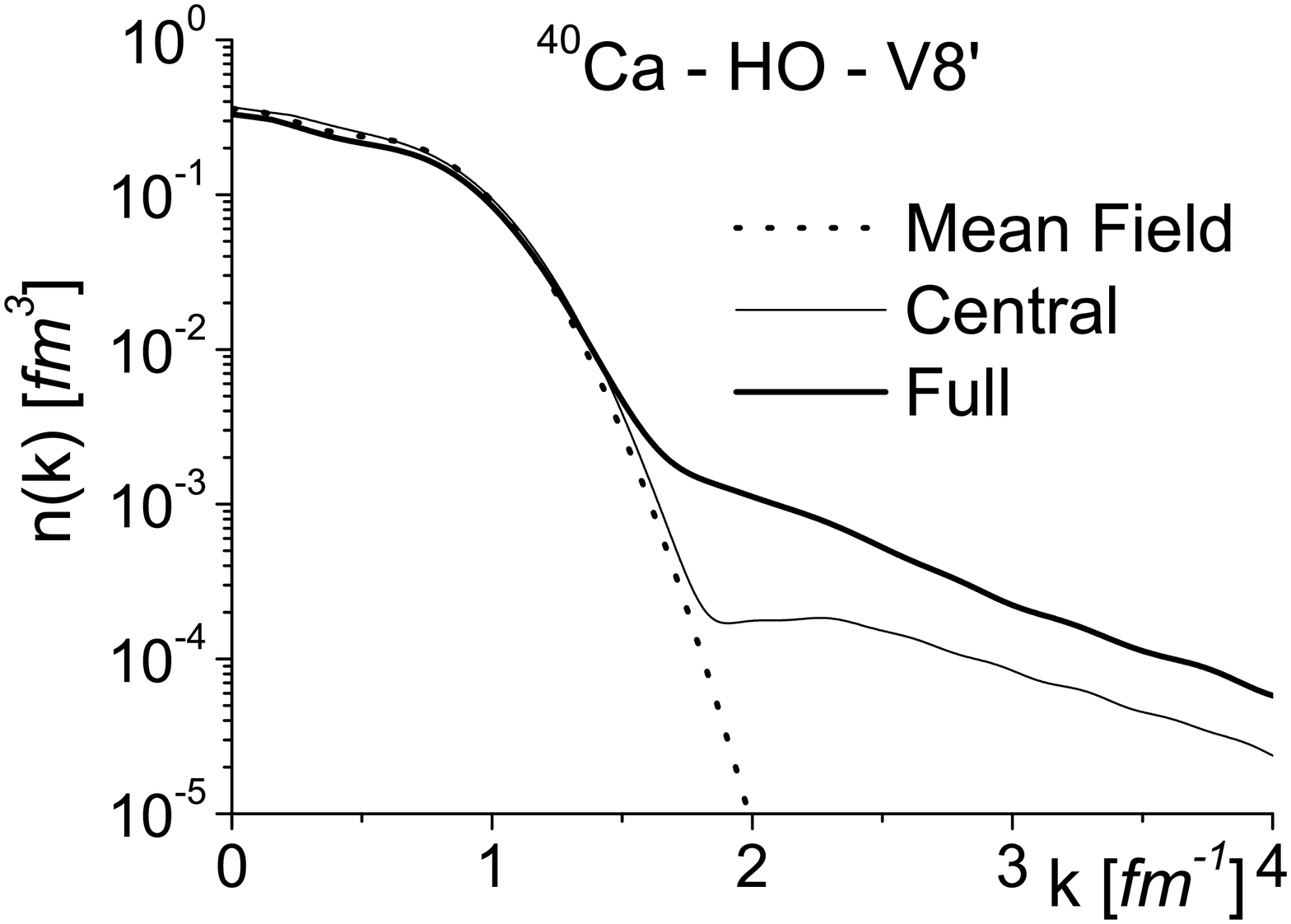}}
  \centerline{\hspace{5mm}
    \epsfysize=0.6\textwidth\epsfbox{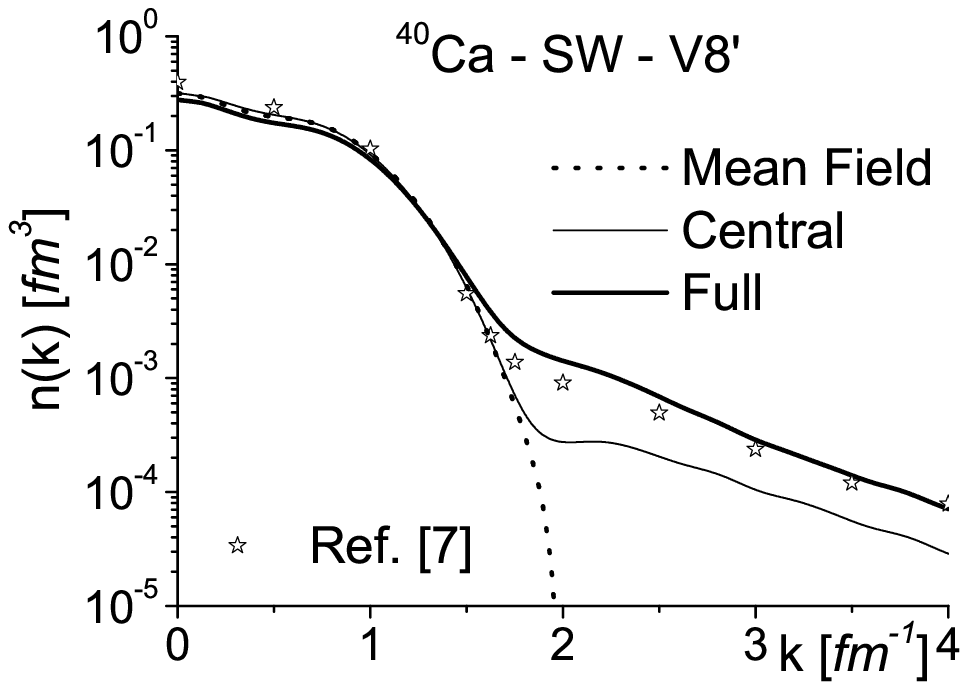}}
  \caption{The same as in Fig. \ref{Fig11}, but for $^{40}Ca$ and
    correlation functions from Fig. \ref{Fig4} and mean field wave functions
    giving the best charge density of Fig. \ref{Fig9}.
     The value of the kinetic energy obtained  by integrating $n(k)$ are
    $\langle T\rangle=$ 782.87 (\textit{central}, HO),
    $\langle T\rangle=$ 1178.45 $MeV$ (\textit{full}, HO);
    $\langle T\rangle=$ 836.24 (\textit{central}, SW) and
    $\langle T\rangle=$ 1245.21 $MeV$ (\textit{full}, SW).}
  \label{Fig12}
\end{figure}

\begin{figure}[!htp]
  \centerline{
    \epsfysize=0.4\textwidth\epsfbox{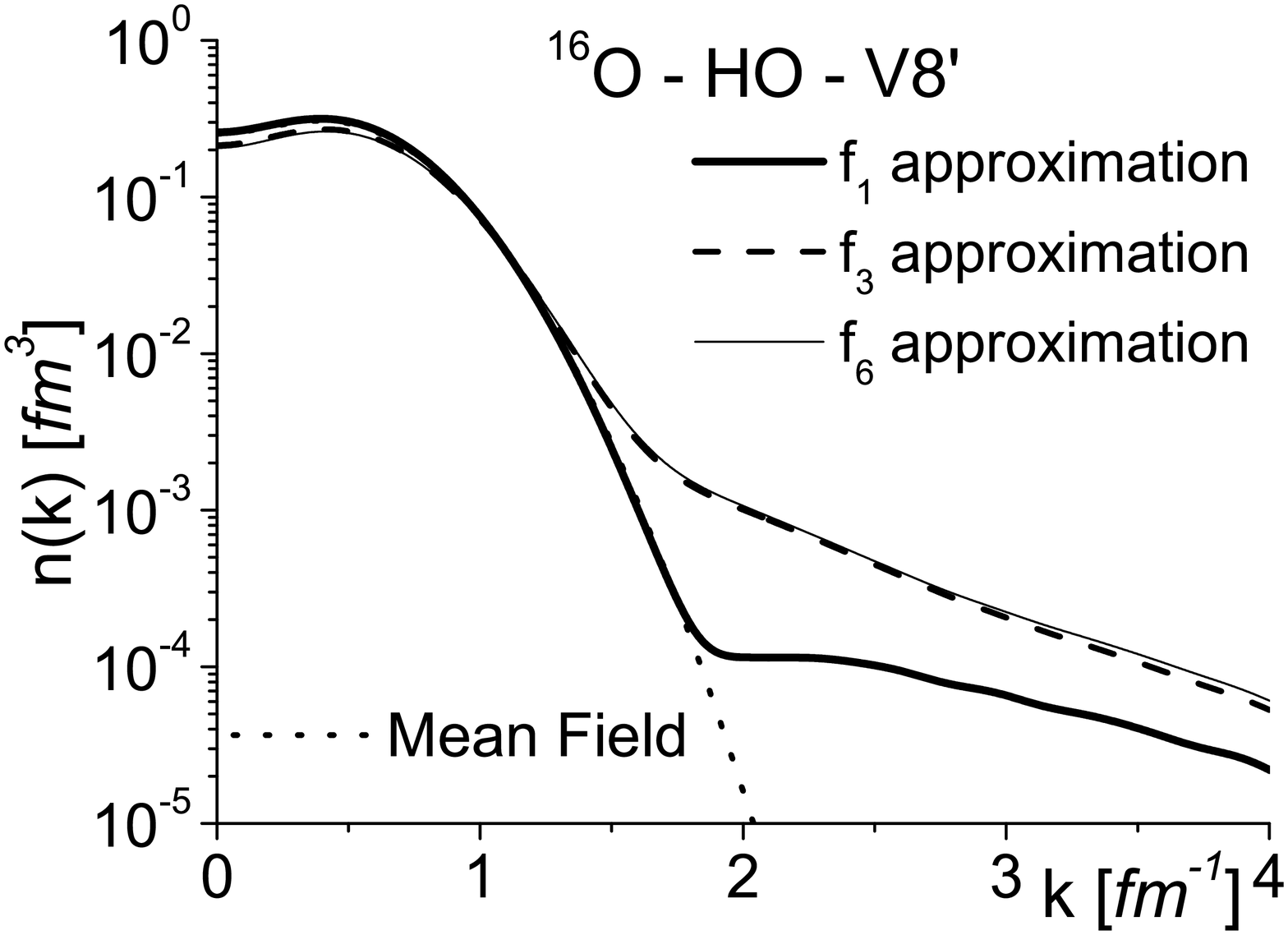}}
  \caption{The effect of the various correlation functions
   on the momentum distribution
    of $^{16}O$. \textit{$f_1$ approximation}: only
    central correlation; \textit{$f_3$ approximation}: $f^{(2)} = f^{(3)} =
     f^{(5)} = 0$.
    \textit{$f_6$ approximation}: full correlation set, $n=1,...,6$.
    Calculations were performed  with correlation functions  taken from  Fig. 3 and HO wave functions.}\label{Fig13}
\end{figure}

\begin{figure}[!hbp]
  \centerline{
    \epsfysize=0.4\textwidth\epsfbox{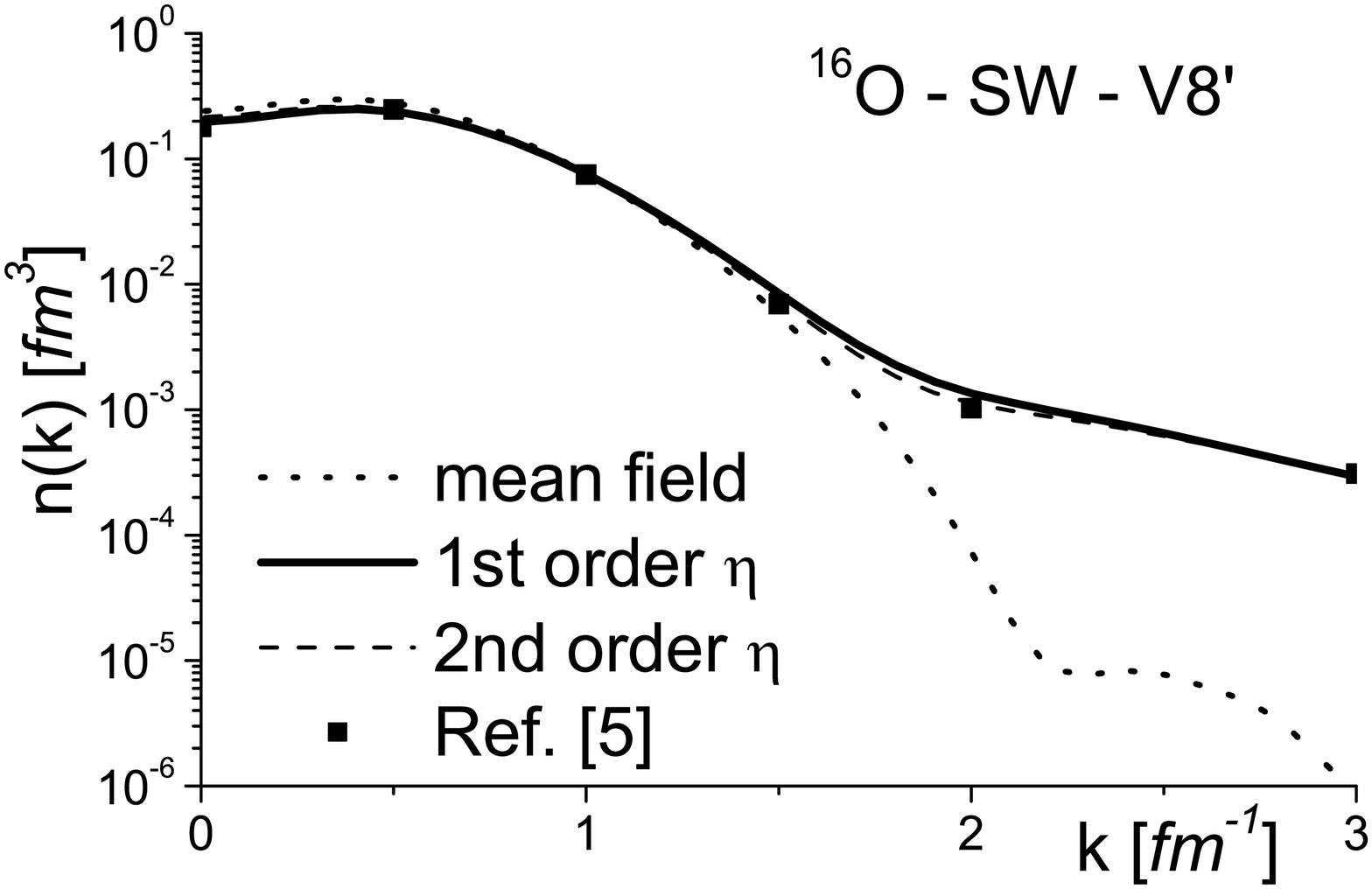}}
  \caption{The convergence of the momentum distributions of $^{16}O$
    calculated by considering the $1st$ and $2nd$ orders of the $\eta$-expansion,
    Saxon-Woods mean field wave functions, the $f_3$ approximation and the correlation functions
    of Fig. \ref{Fig3}. Our results are compared with the results of Ref. \cite{pie01}
    obtained within the VMC approach and the AV14 interaction.
    The values of the kinetic energies obtained by integrating the
    momentum distributions are: $\langle T\rangle=$ $521.87$ $MeV$ (mean field),
    $\langle T\rangle=$ $980.10$ $MeV$ ($1st$ order $\eta$),
    $\langle T\rangle=$ $932.64$ $MeV$ ($2nd$ order $\eta$).
    The normalization of $n(k)$ is $\int d\Vec{k}n(k) = 1$.}
  \label{Fig14}
\end{figure}

Using the correlation functions shown in Figs. \ref{Fig3} and \ref{Fig4},
and the mean field WF's corresponding
to the best densities shown in Figs. \ref{Fig8}  and  \ref{Fig9} , we have
calculated the momentum distributions given by Eq. (\ref{defmomdis}), with
the OBMD matrix $\rho^{(1)}(\Vec{r}_1,\Vec{r}_1^\prime)$ given by Eq. (\ref{obrhoeta}).
The results, obtained  at first order in $\eta$ (the convergence will be discussed
later on), are presented in Figs. \ref{Fig11}, \ref{Fig12}, \ref{Fig13} and
\ref{Fig14}. In Figs. \ref{Fig11} and \ref{Fig12}, our results are compared
 with the the results obtained
in Ref. \cite{fab01}, where the same interaction and the same
correlation functions have been used, and in case of $^{16}O$, also
with the results of Ref. \cite{pie01}
 where  the $AV14$ interaction \cite{pud01} and the Variational Monte Carlo
  approach have been used.
 These comparisons show that:
 \begin{enumerate}
 \item our results nicely agree with the ones of Refs. \cite{fab01} and \cite{pie01};
 \item
short range central correlations do not produce enough high
        momentum components, although they appreciably affect the
      momentum distributions at $k\ge 2$ $fm^{-1}$;the inclusion of the tensor operators greatly enhances the
        high-momentum tail of the distribution in the region  $k> 2$
        $fm^{-1}$.

 \item   the largest effect from non central correlations
      comes from tensor ($\hat{S}_{ij}$) and tensor-isospin
      ($\hat{S}_{ij}\,\Vec{\tau}_i\cdot\Vec{\tau}_j$) correlations
      ($n=4$ and $n=6$ in Eq. (\ref{newuno})), the other components
      playing a minor role; thus, as shown in Fig.  \ref{Fig13}, the $f_3$
      approximation appears to be a rather good one
      for the calculation of the momentum distributions;

\item[iv)]{the satisfactory agreement of our results with the ones of Refs.
        \cite{fab01} and \cite{pie01} shows that the convergence of the
      $\eta$-expansion for $\rho(\Vec{r},\Vec{r}^\prime)$ is a very good one.
      As a matter of fact, we have explicitly evaluated the next order cluster
      contribution for $^{16}O$; the results, reported in Fig. \ref{Fig14}
        using the $f_3$ approximation for the correlations functions,
        show a very good convergence indeed.}
\end{enumerate}

We would like to stress that the good convergence of the momentum
distributions is a proof of the good convergence
of $\langle T\rangle$.

%%%%%%%%%%%%%%%%%%%%%%%%%%%%%%%%%%%%%%%%%%%%%%%%%%%%%%%%%%%%%%%%%%%%%%%%%%%%%%%%%%%%%%%%%%%

\section{Effects of correlations on the charge density and momentum distributions:
the diagrammatic description}\label{section7}

\begin{figure}[!hp]
\centerline{\textit{a)}\hskip 1cm
      \epsfysize=0.9cm\epsfbox{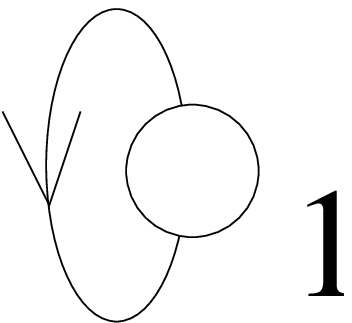}}
\vskip 0.8cm
\centerline{\textit{b)}\hskip 1cm
      \epsfysize=1.cm\epsfbox{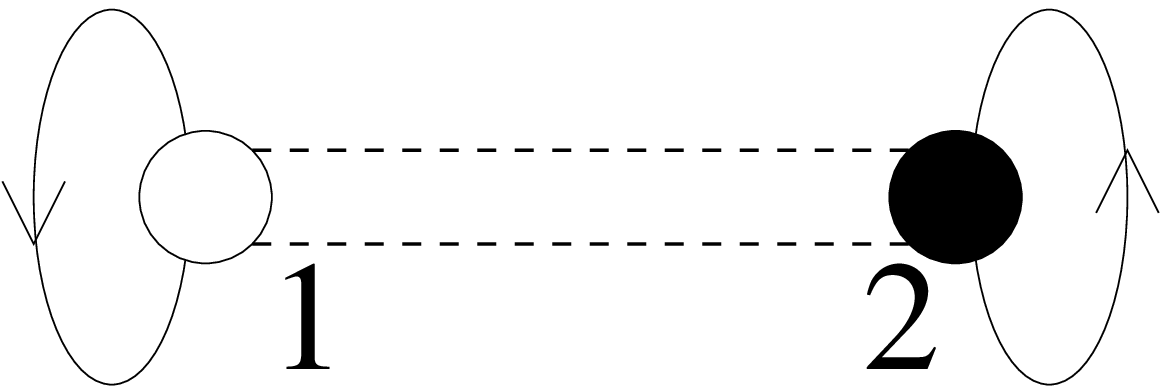}
      \hskip 2.5cm
      \epsfysize=1.1cm\epsfbox{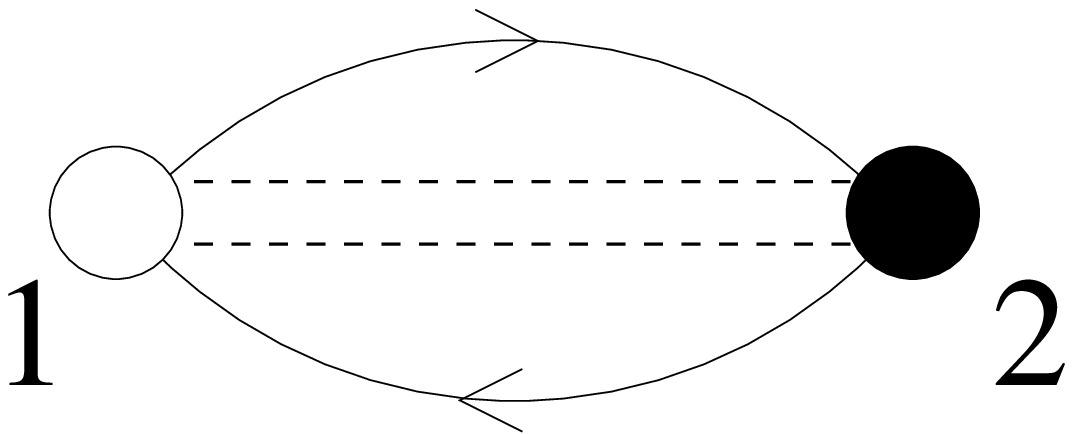}}
\vskip 1cm
\centerline{\textit{c)}\hskip 1cm
      \epsfysize=2.5cm\epsfbox{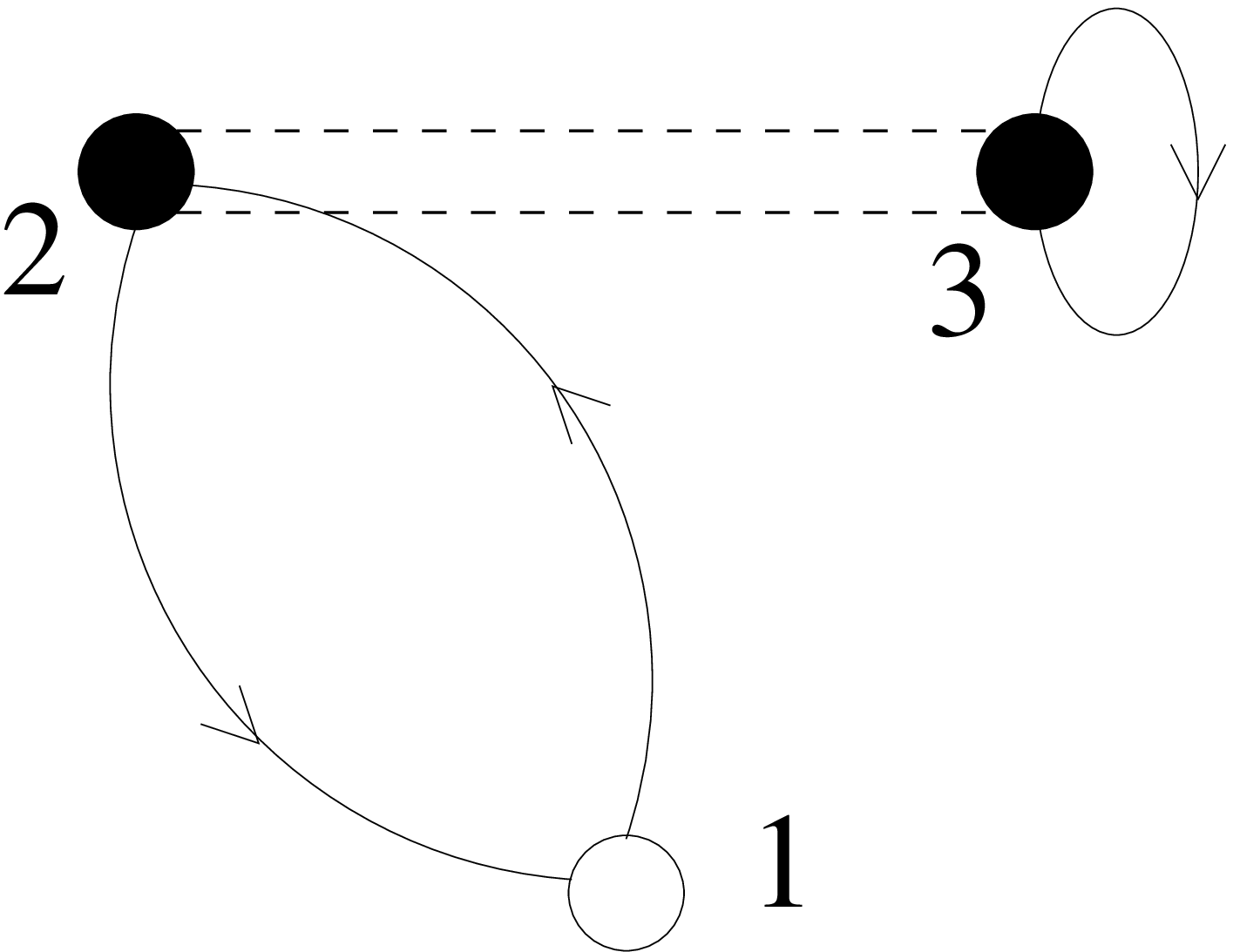}
      \hskip 3cm
      \epsfysize=2.5cm\epsfbox{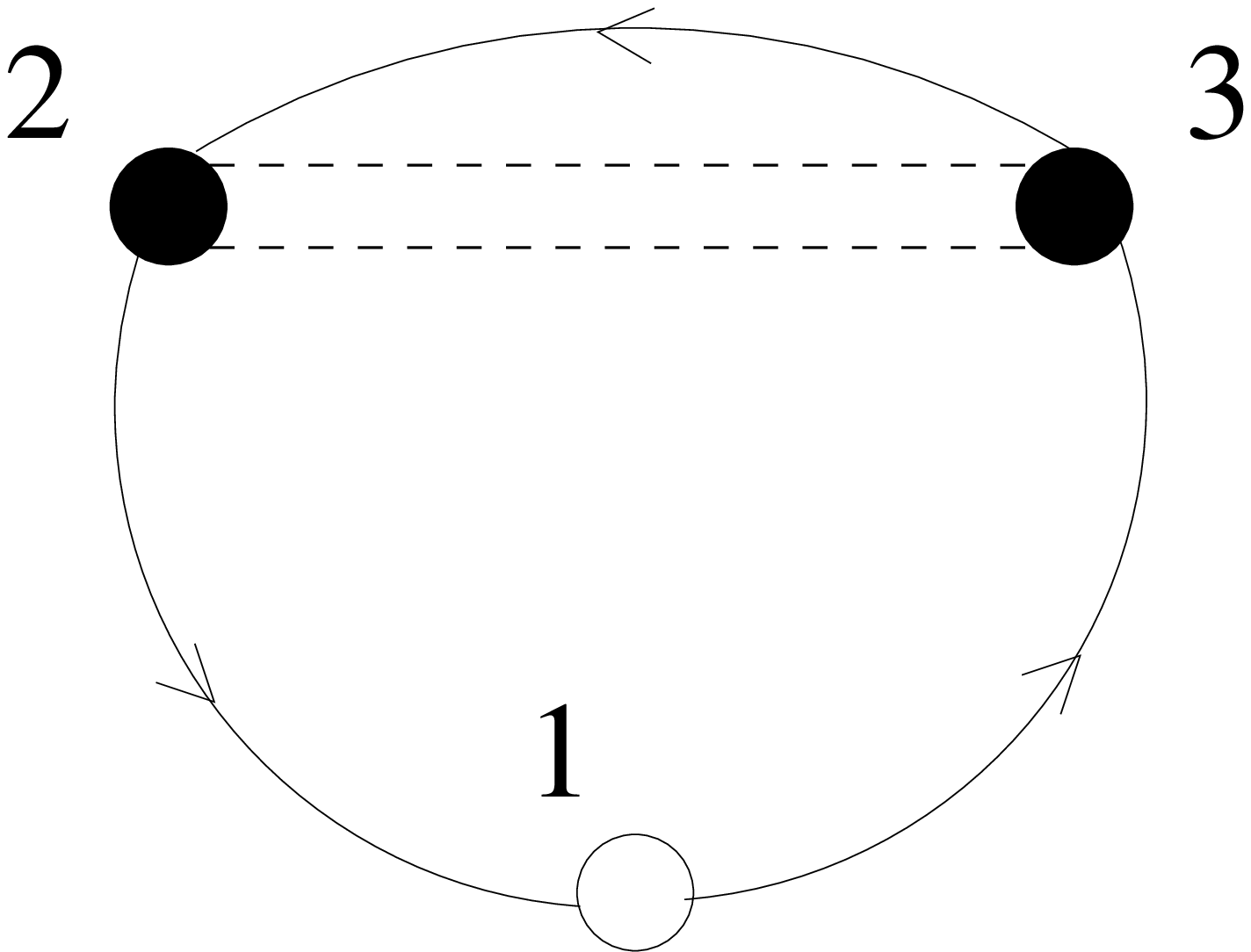}}
\caption{Diagrammatic representation of the one body density $\rho(\Vec{r}_1)$
      in the lowest order
      of the $\eta$-expansion (Eq. (\ref{rhodens})). Dots denote
      the corresponding spatial coordinates and integrations occur over full
      dots. An oriented full line represents the shell model uncorrelated
      non diagonal density matrix $\rho^{(1)}_o(i,j)$, an oriented closed line
      the diagonal density matrix $\rho_o(i,i)$, and a dotted line
      the correlation $H(r_{ij},r_{ji})$ between particles $i$ and $j$.
      \textit{a)}: shell model, uncorrelated density $\rho_o$;
      \textit{b)}: hole contribution $\Delta\rho^H$
      \textit{c)}: spectator contributions $\Delta\rho^S$.
      The direct and exchange contributions are shown on the left and right sides
      of the Figure, respectively.}\label{Fig15}
\vskip -0.1cm
\end{figure}

\begin{figure}[!hp]
\centerline{\textit{a)}\hskip 1cm
      \epsfysize=0.6cm\epsfbox{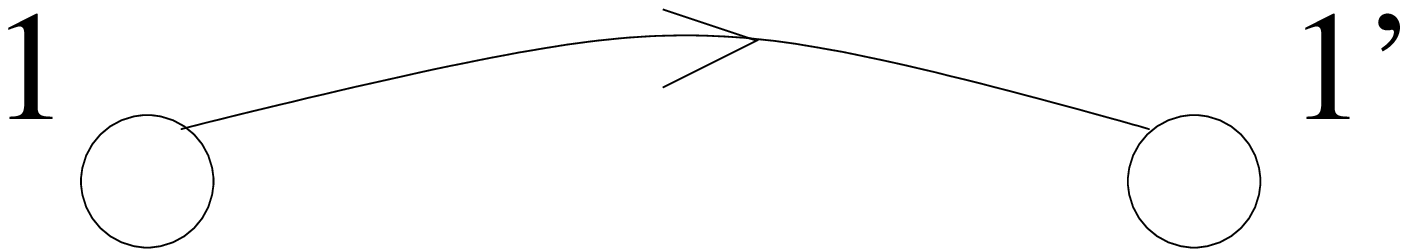}}
\vskip 0.8cm
\centerline{\textit{b)}\hskip 1cm
      \epsfysize=2.5cm\epsfbox{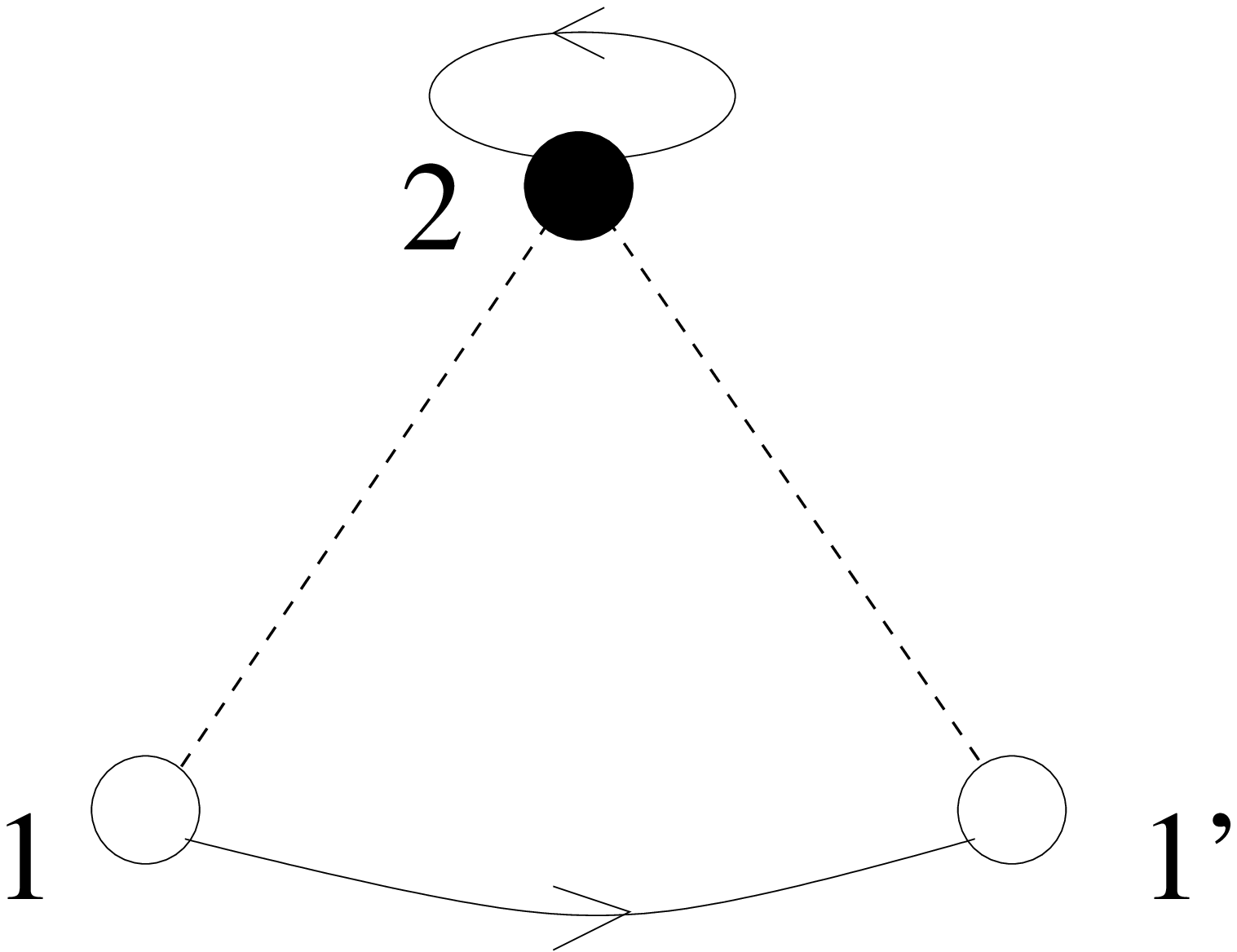}
      \hskip 1cm
      \epsfysize=2.5cm\epsfbox{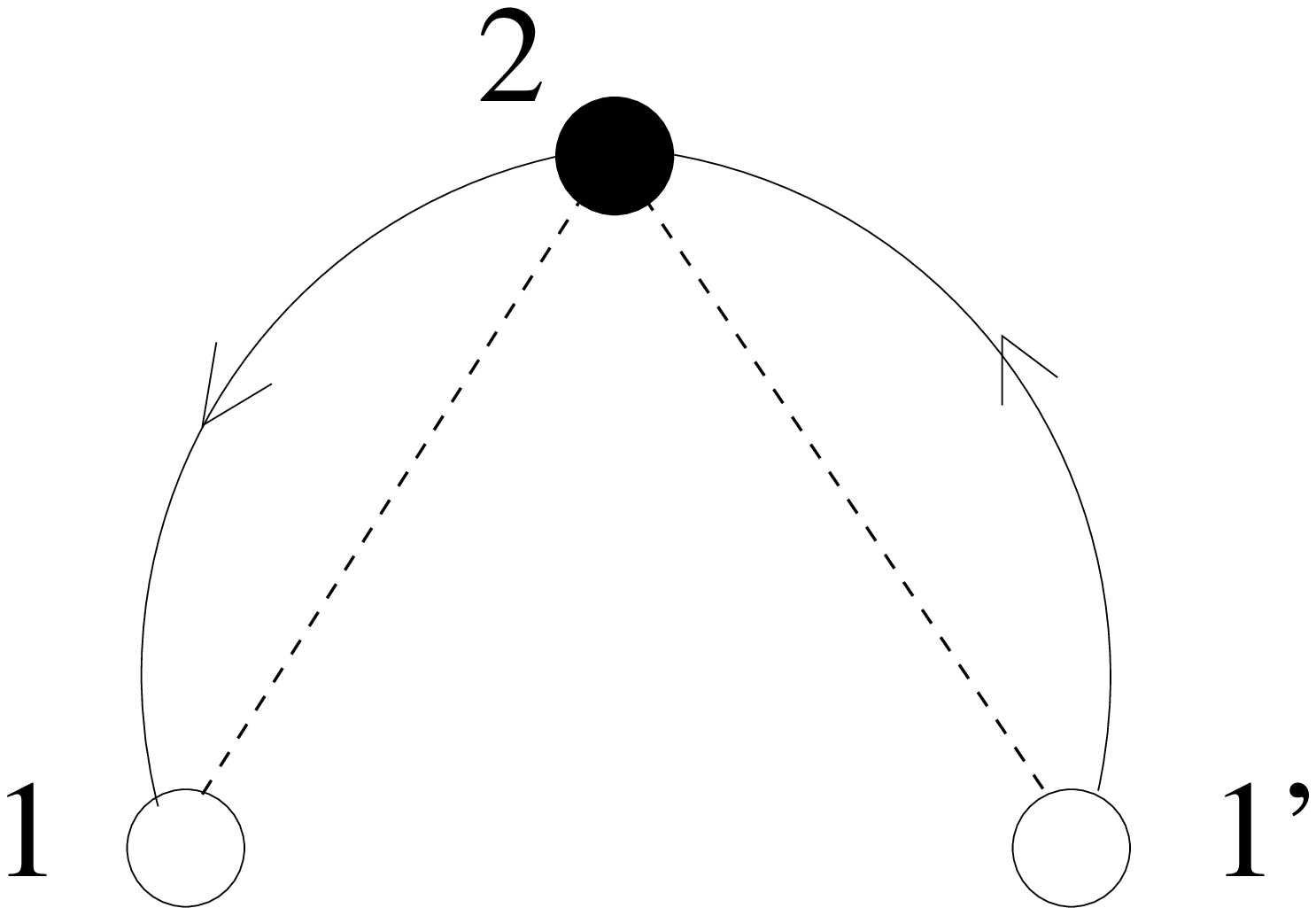}}
\vskip 1cm
\centerline{\textit{c)}\hskip 1cm
      \epsfysize=2.5cm\epsfbox{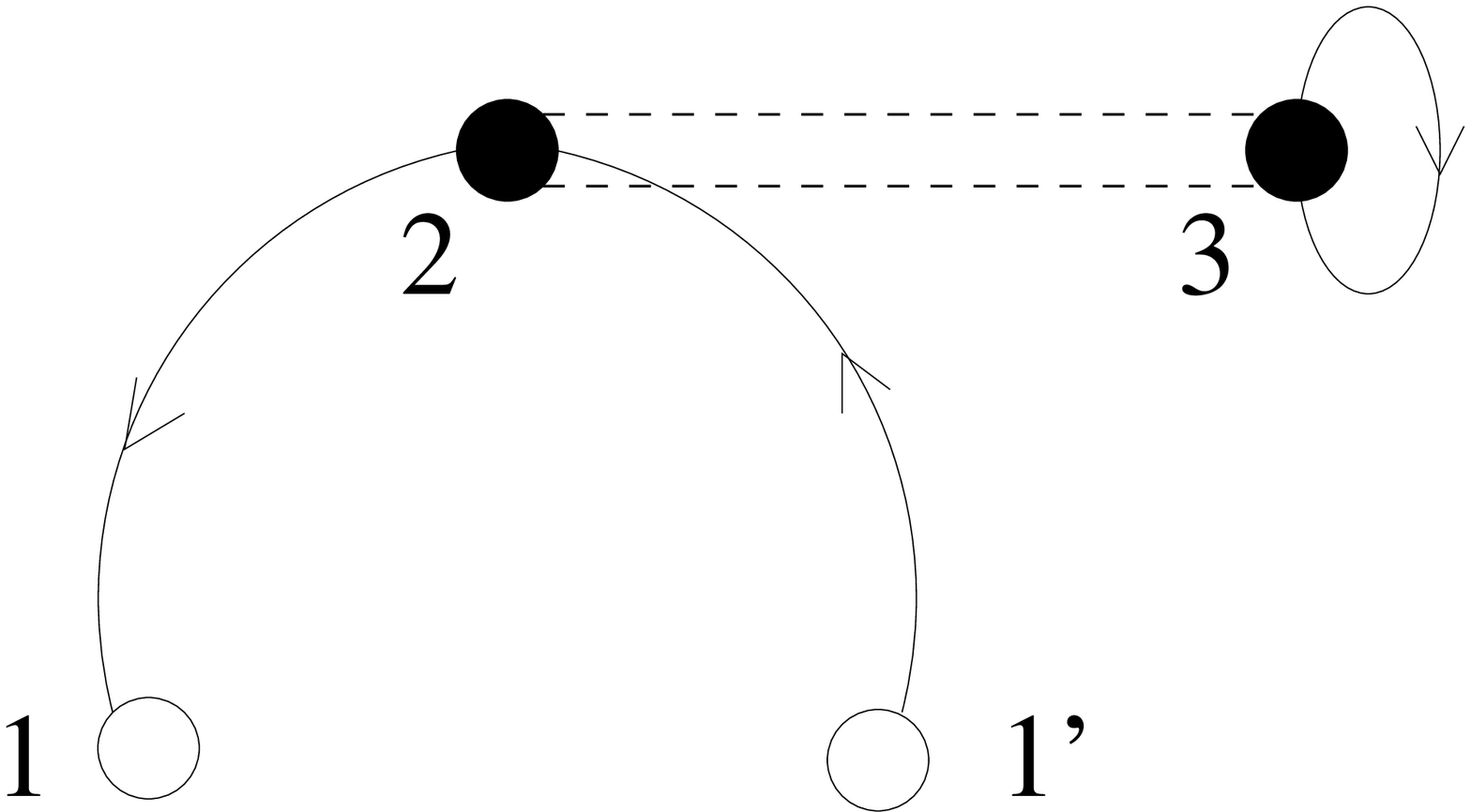}
      \hskip 0.5cm
      \epsfysize=2.5cm\epsfbox{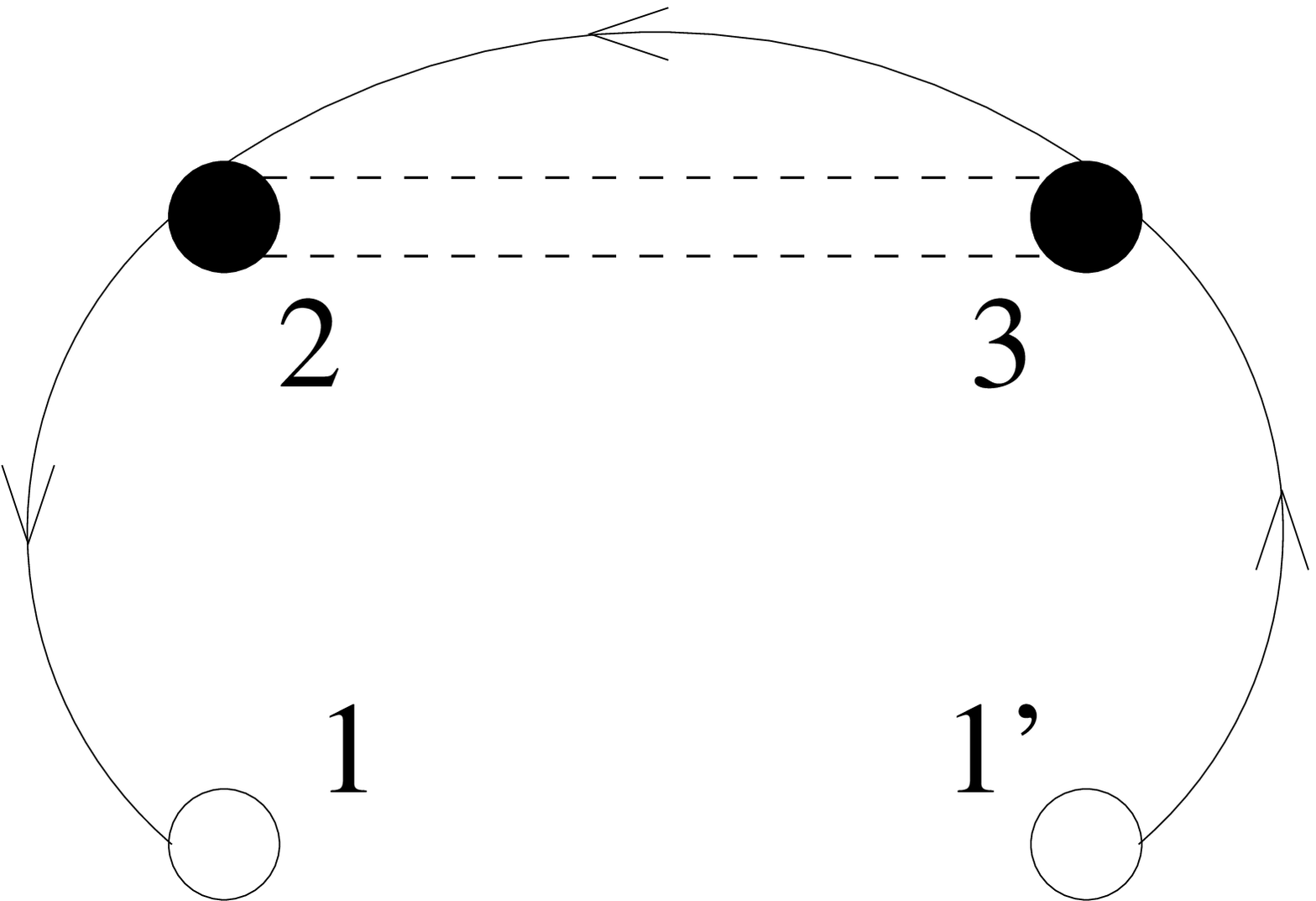}}
\caption{The same as in Fig. \ref{Fig15} for the one body mixed density matrix
  $\rho^{(1)}(\Vec{r},\Vec{r}^\prime)$.}\label{Fig16}
\vskip -0.1cm
\end{figure}

\begin{figure}[!htp]
  \centerline{\epsfxsize=0.7\textwidth
    \epsfbox{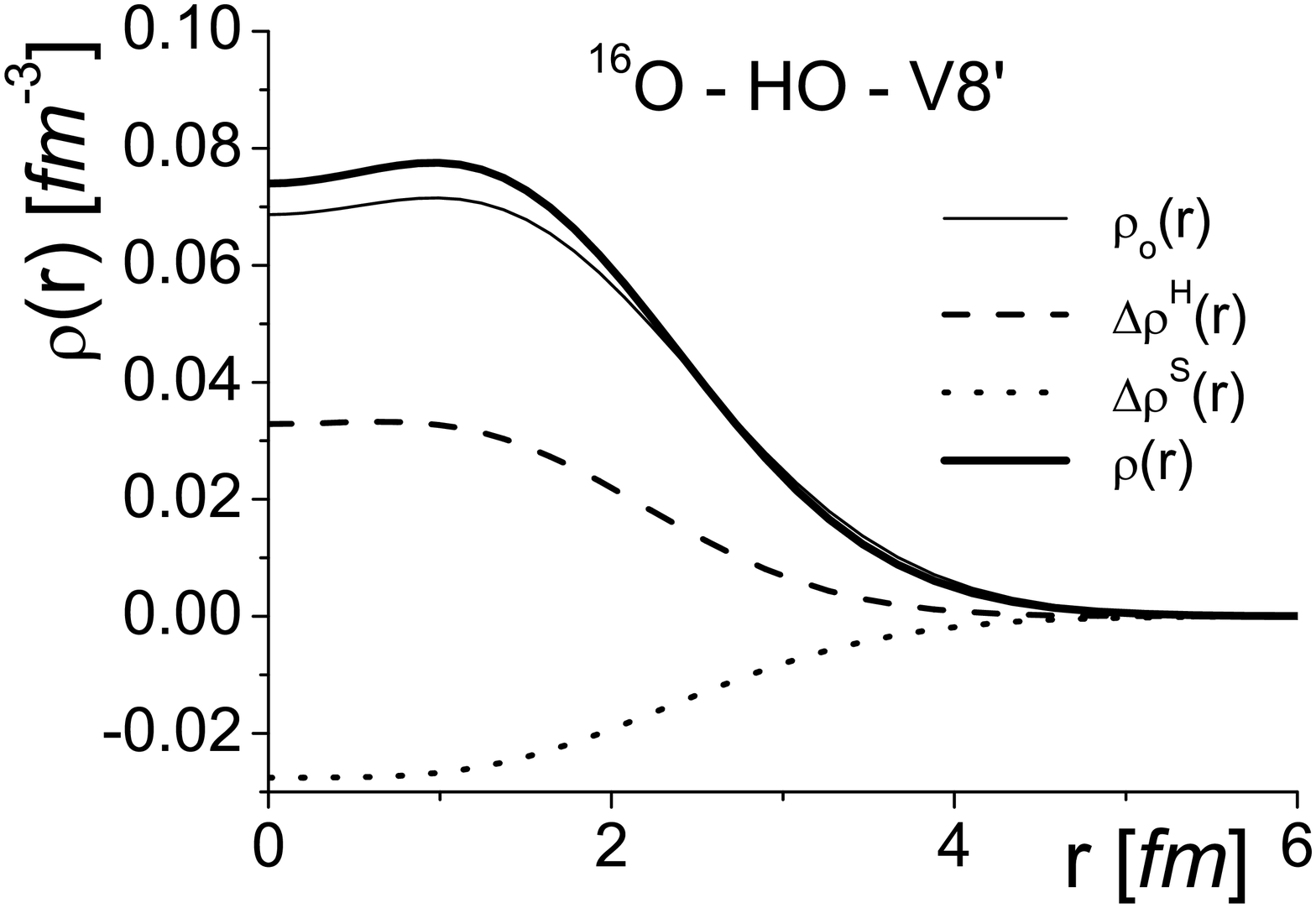}}
  \caption{The charge density of $^{16}O$.
      \textit{Thin full}: shell model density (Eq. (\ref{rhoSM})).
      \textit{Thick full}: correlated density (Eq. (\ref{rhodens}))
      calculated with the correlation functions of Fig. \ref{Fig3} \cite{fab01}
      and HO mean field Wave Functions with parameter $a=1.8$ $fm$.
      \textit{Dashed}: hole contribution $\Delta\rho^H(r)$;
      \textit{dotted}: spectator contribution $\Delta\rho^S(r)$.}
  \label{Fig17}
\end{figure}

\begin{figure}[!hbp]
  \centerline{\epsfxsize=0.7\textwidth\epsfbox{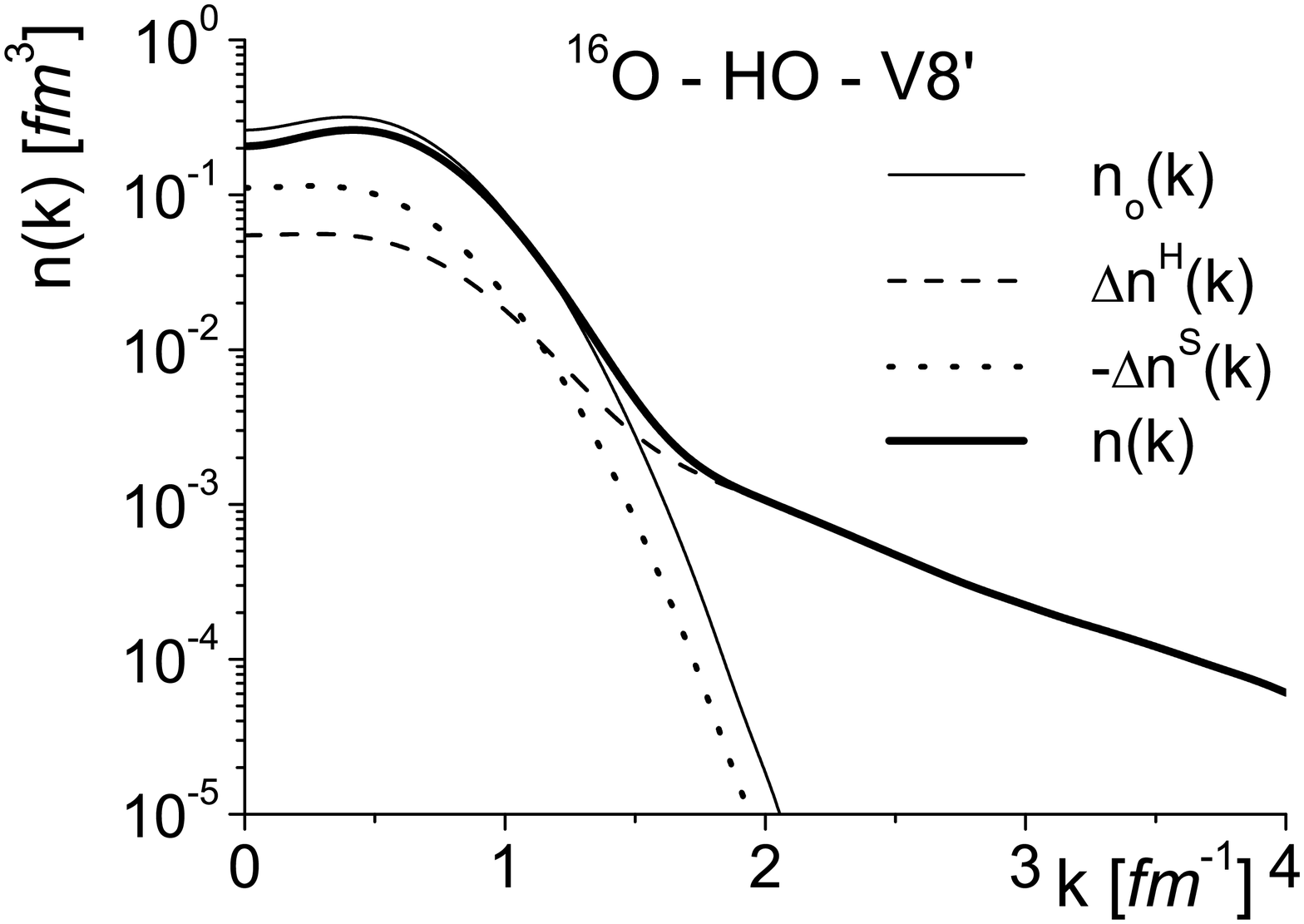}}
  \caption{The same as in Fig. \ref{Fig17}, but for the momentum distribution.}
  \label{Fig18}
\end{figure}

 Within any body approach based upon
the correlated wave function  ($\ref{psi1}$), the interaction ($\ref{defpotere}$)
and the cluster expansion technique, any quantity
 can be described  by a
 diagrammatic representation, which provides  a meaningful definition of correlations
  and
the extent to which they affect the given quantity. In particular,
 the density and the momentum distributions are associated to diagrams
 according to the following rules (see e.g. Ref. \cite{gau01}):
 \begin{enumerate}
 \item  an open dot with the index
$i$ denotes the coordinate $\Vec{r}_i$;
\item  a full dot with index $i$ stands for
integration over $\Vec{r}_i$;
\item  an oriented line joining two dots  with indexes $i$ and $j$
denotes $\rho^{(1)}_o(\Vec{r}_i,\Vec{r}_j)$;
\item a line beginning from and ending
in a dot with index $i$ denotes $\rho_o(\Vec{r}_i)$;
\item a dashed line joining two full 
dots with indexes
$i$ and $j$ denotes $H^{D(E)}(\Vec{r}_{ij})$;
\item two dashed lines joining
the open dots with index $i$ and $j$ with the full dot denotes
$H^{D(E)}(\Vec{r}_{ij},\Vec{r}_{i^\prime j})$.
\end{enumerate}
The diagrammatic representations  of the diagonal (Eq.
(\ref{rhodens})) and non diagonal (Eq. (\ref{obrhoeta})) one body
density matrices are shown in Figs. \ref{Fig15} and \ref{Fig16}
respectively. The meaning of \textit{hole} ($H$) and
\textit{spectator} ($S$) contributions introduced in Section
\ref{sectionIVA} (cf. Eqns. (\ref{39}) and (\ref{40}))becomes now
clear: the first, represented by
 diagrams $15b$ ($16b$)
 describes the process in which particle
$''1''$ is correlated with particle $''2''$,  whereas the second one, represented
by diagrams  diagrams $15c$ ($16c$), describes
the process in which dynamical correlations are acting between particles
$''2''$ and `$''3''$. In Figs. \ref{Fig17} and \ref{Fig18} we show the effect
of the hole and spectator contributions  on the charge density and momentum distributions,
respectively. It can be seen that the effects on the two quantities are very different:
as far as the density is concerned, $\Delta\rho^H$ and $\Delta\rho^S$ are almost
 of the same value,
and of opposite sign,  with a small net effect; as for the momentum
distribution, the spectator contribution only affects the SM
distribution by an almost constant factor of $0.8$, whereas the hole
contribution create the large amount of high momentum components.
The spectator contribution leads to a renormalization of the mean
field orbitals and to a decrease of the occupation number for states
below the Fermi level, whereas the hole contribution is responsible
for the high momentum components. This explains and qualitatively
justifies the parameterized $n(k)$ of Ref. \cite{ciofisimula} in the
form $n(k) = n_o(k) + n_1(k)$. It is clear, that the amount of hole
and spectator
 correlations also
depends upon the amount of the mean field contributions;  calculations show however
that the latter, even if obtained within the most sophisticated
mean field approaches,  cannot never provide, e.g. the amount of high momentum components
generated by the hole contribution, so that the high momentum part of $n(k)$ is
practically  due   only to (hole) correlations. Other quantities which are very sensitive
to hole correlations will be discussed elsewhere  \cite{isabella}.

\section{Summary and conclusions}

In this paper we have addressed the problem of developing a method which
could be used to calculate scattering processes at medium and high energies
within a realistic and parameter-free  description of nuclear structure.
The $\eta$-expansion
seems to satisfy such a requirement: as a matter of fact, it can be used within the following
strategy: \textit{i)} the values of the parameters pertaining to the correlation functions and
the mean field wave functions, can be obtained from the calculation of the
ground-state
energy, radius and density of the nucleus,  to a given order of the expansion;
\textit{ii)} using these parameters, any scattering process can be evaluated at the same order of
the cluster expansion.
 The method therefore appears to be a
very effective,  transparent and parameter-free one.
It should however be pointed out that, as
any other many body approach, our cluster expansion approach  may suffer from the well
known convergence problem, so that  the role played by the disregarded higher order
terms has  to be estimated. This is precisely what has been done
  in the present paper, adopting  the following
procedure: \textit{i)} our lowest order results have  been compared
with the ones obtained within more complete approaches,  like \eg
the FHNC and VMC methods, and \textit{ii)} a direct calculation of
the  higher order terms of the momentum distribution $n(\Vec{k})$
has been performed. It turned out that  the value of the
ground-state energy calculated within the first order
$\eta$-expansion  reasonably agrees  with the one obtained
  within the FHNC/SOC approach. The agreement is very good as far as the average
  kinetic energy is concerned,  whereas  differences occur in some of
  the potential energy contributions, as it should have been expected due to
  the complex spatial dependence of some of the components of the nucleon-nucleon
  interaction. Nonetheless, using the same correlation functions as in the FHNC/SOC
  calculation, we obtain a reasonable minimum value of the energy, with mean field WF
  very near to the ones of the FHNC/SOC approach. Furthermore,
  our results for the charge
  density and momentum distributions shows a very good agreement with the
  results obtained within the FHNC/SOC approach and even with the VMC approach, and
   the direct calculation of the higher order terms in the expansion
    of the momentum distributions shows a very good convergence of the $\eta$-expansion
    up to very high values of the momentum.

To sum up, we have shown that, using realistic models of the nucleon-nucleon interaction,
 a proper approach based on cluster expansion techniques  can produce reliable approximations
for those diagonal and non diagonal density matrices which appear in various
medium and high energy scattering processes off nuclei, so that the role of nuclear effects
in these processes can be reliably estimated without using free parameters to be fitted
to the data.  The approach has already been extended
to the treatment of  the final state interaction effects in
$A(e,e'p)X$ processes at medium energies within the eikonal-Glauber multiple
scattering theory,
and to the calculation of  nuclear and color transparencies effects. Preliminary
 results \cite{alv02} are very encouraging. Calculations of other types of high energy
scattering processes (e.g.the total nucleon-Nucleus cross section)
 are in progress and will be reported elsewhere \cite{isabella}.

\section{Acknowledgments}
We are grateful to Giampaolo Co' for providing the FHNC results used in the
benchmark calculation and to Adelchi Fabrocini who supplied us with the correlation
functions corresponding to the $V8'$ interaction. One of us (CcdA) would like to thank Rafael Guardiola 
for enlightening discussions.

%%%%%%%%%%%%%%%%%%%%%%%%%%%%%%%%%%%%%%%%%%%%%%%%%%%%%%%%%%%%%%%%%%%%%%%%%%%%%%%%%%%%%%%%%%%

\appendix

\section{Mean field wave functions}

The mean field wave functions  have the general form:
\beq
\psi_{nlm}(\Vec{r})\,=\,R_{nl}(r)\,Y_{lm}(\theta,\varphi)
\eeq
with $R_{nl}(r)$ the radial part and $Y_{lm}(\theta,\varphi)$ the spherical harmonics.
We have used harmonic oscillator and Saxon-Woods wells to generate  the  radial part;
 in the harmonic oscillator case, we have
\beq
\label{eqHOwf}
R_{nl}(r)\,=\,e^{-x/2}\,x^{l/2}\,U_{nl}\,X_{nl}^{1/2}\Psi
 \eeq
  with $x=r^2/a^2$, $a$ is the HO parameter and
\beq U_{nl}\,=\,\sum_{k=1}^n\,\frac{(-1)^k\,x^k\,n!\,2^k\,(2l+1)!!}
                              {(n-k)!\,k!\,(2l+2k+1)!!}\,,
\eeq
\beq
X_{nl}\,=\,\frac{2^{l-n+2}\,(2l+2n+1)!!}
                {{(2l+1)!!}^2\,n!\,\pi^{1/2}\,a^3}\,;
\eeq
whereas, in the Saxon-Woods  case,  the radial
part is the solution of the radial Schr\"odinger equation with one-body potential
of the following form
\beq
V_{SW}(r)\,=\,-\,\frac{V_o}{1+e^{-(r-R_o)/a_o}}\,.
\eeq

\section{Parameterization of the correlation functions for $^{16}O$ and $^{40}Ca$}

The correlation functions $f^{(n)}(r)$ for $^{16}O$ and $^{40}Ca$, shown in Figs. \ref{Fig3}
and \ref{Fig4}, respectively, can be conveniently parameterized in the following way:
\beqy\sublabon{equation}
f^{(n)}(r)&=&\sum_{i=1}^6\,A^{(n)}_i\,e^{-\,B^{(n)}_i\,r^i}\hspace{1.5cm}n=1,...,5
\label{o16corfuncA}\\
f^{(6)}(r)&=&A^{(6)}_2 r^2 e^{-B^{(6)}_1 r}\,+\,\sum_{i=2}^6\,A^{(6)}_i\,r^{i-1}
                \,e^{-\,B^{(6)}_i\,r^i}
                \label{o16corfuncB}
\eeqy\sublaboff{equation}
where the parameters $A^{(n)}_i$ and $B^{(n)}_i$ are given in Table \ref{Table9}
for $^{16}O$ and in Table \ref{Table10} for $^{40}Ca$.

\section{The coefficients of the one body non diagonal density matrix
resulting from the spin-isospin summation}

The coefficients appearing in Eq. (\ref{accadirex}) for the OBMD
are defined as \beqy C^{(p,q)}_{D}(r_{12},r_{1^\prime 2})&=&
\sum_{\sigma_1,\sigma_2,\tau_1,\tau_2}
 \bra{\sigma_1\tau_1\,\sigma_2\tau_2}
       \,\hat{O}^{(p)}_{12}\,\hat{O}^{(q)}_{1\prime 2}\,\ket{\sigma_1\tau_1\,\sigma_2\tau_2}\,,\nn\\
\,C^{(p,q)}_{E}(r_{12},r_{1^\prime2})&=&
\sum_{\sigma_1,\sigma_2,\tau_1,\tau_2}
\bra{\sigma_1\tau_1\,\sigma_2\tau_2}
       \,\hat{O}^{(p)}_{12}\,\hat{O}^{(q)}_{1^\prime 2}\,\ket{\sigma_2\tau_2\,\sigma_1\tau_1}
\eeqy and can be calculated analytically;  their  explicit values are
summarized in Table \ref{TableMatrix}.

The coefficients appearing in the definition of the TBD 
are more involved and can only be written in terms of the
spin-isospin states upon which they have to be calculated:
\beqy
A^{(n=1,6)}_D&=&\{16,0,0,0,0,0\}\,;\hspace{3cm}A^{(n=1,6)}_E\,=\,\{4,12,12,36,0,0\}\,;\\
B^{(p,q)}_{(n),\mathcal P} &=&
 \sum_{\sigma,\tau}\,  \bra{\sigma_1\tau_1,\,\sigma_2\tau_2,\,\sigma_3\tau_3}
      \,\hat{O}^{(p)}_{13}\,\hat{O}^{(n)}_{12}\,\hat{O}^{(q)}_{13}
      \,\ket{\sigma_1\tau_1,\,\sigma_2\tau_2,\,\sigma_3\tau_3)}_{\mathcal P}\\
C^{(p,q)}_{(n),\mathcal P}&=&
\sum_{\sigma,\tau}\,\bra{\sigma_1\tau_1,\,\sigma_2\tau_2,\,\sigma_3\tau_3}
      \,\hat{O}^{(p)}_{23}\,\hat{O}^{(n)}_{12}\,\hat{O}^{(q)}_{23}
      \,\ket{(\sigma_1\tau_1,\,\sigma_2\tau_2,\,\sigma_3\tau_3)}_\mathcal{P}\\
D^{(q)}_{(n),\mathcal P} &=&
\sum_{\sigma,\tau}\,\bra{\sigma_1\tau_1,\,\sigma_2\tau_2,\,\sigma_3\tau_3,\,\sigma_4\tau_4}
      \,\hat{O}^{(n)}_{12}\,\hat{O}^{(q)}_{34}
      \,\ket{(\sigma_1\tau_1,\,\sigma_2\tau_2,\,\sigma_3\tau_3,
      \,\sigma_4\tau_4)}_\mathcal{P}
\eeqy
where only \textit{linked}
permutations are considered; for example, in the four-body term, the identical permutation
$\ket{\alpha_1\beta_2\gamma_3\delta_4}$ is not linked, because the only present links
are between particles $12$ and $34$, but the two clusters are not connected; there
are four unlinked permutations in this term.

%%%%%%%%%%%%%%%%%%%%%%%%%%%%%%%%%%%%%%%%%%%%%%%%%%%%%%%%%%%%%%%%%%%%%%%%%%%%%%%%%%%%%%%%%%%

%---------------------------------------------------------------------Table I

%---------------------------------------------------------------------Table I
\newpage
\begin{table}[!hp]
  \caption{The results of the benchmark calculation of the ground-state energy of
  $^{16}O$ using the $V8^\prime$ interaction \cite{pud01}, the correlation
  functions shown in Fig. \ref{Fig1} \cite{30a}, and harmonic oscillator
  mean field wave functions with parameter $a=2$ $fm$ (cf. Appendix A1).
  The results of  the $\eta$-expansion
  obtained in this paper are compared  with the FHNC/SOC results of Ref. \cite{30a}.
  $\langle V\rangle$ is the average potential energy, $\langle T\rangle$ the average
  kinetic energy, $E=\langle V\rangle+\langle T\rangle$ the total energy,
  and $E/A$ the total energy per particle. The kinetic energy of the Center-of-Mass
  motion has been subtracted from the expectation value of
  the kinetic energy operator. All quantities in $MeV$.}
    \begin{center}
      \begin{tabular}{l||c|c|c|c}
approach&$\langle V\rangle$&$\langle T\rangle$&$E$&$E/A$\\\hline
$\eta$-expansion, \textit{this paper}
                 & -390.37 & 323.50 & -65.90 & -4.12\\
\textit{FHNC/SOC}, Ref. \cite{30a}
                 & -390.30 & 325.18 & -65.12 & -4.07\\\hline
      \end{tabular}
   \end{center}
\label{Table1}
\end{table}
%--------------------------------------------------------------------Table II
%\newpage
\begin{table}[!hp]
   \caption{The contributions $\langle V_i\rangle$ of the first six channels of
     the $V8^\prime$ interaction to the average potential energy $\langle V\rangle$
     shown in Table \ref{Table1}.
     All quantities in $MeV$.}
   \begin{center}
     \begin{tabular}{l||c|c|c|c|c|c|c}
approach&$\langle V_{c}\rangle$&$\langle V_{\sigma}\rangle$&$\langle V_{\tau}\rangle$
        &$\langle V_{\sigma\tau}\rangle$&$\langle V_{t}\rangle$&$\langle V_{t\tau}\rangle$
        &$V=\sum_{i} V_i$\\\hline
$\eta$-expansion, \textit{this paper}
                 & 0.6 & -35.4 & -10.1 & -172.8 &-0.03& -172.7 & -390.37\\
\textit{FHNC/SOC}, Ref. \cite{30a}
                 & 0.7 & -40.1 &-10.6 & -180.0 & 0.07 & -160.3 & -390.30\\\hline
      \end{tabular}
   \end{center}
\label{Table2}
\end{table}
%---------------------------------------------------------------------Table III
\newpage
\begin{table}[!hp]
  \caption{The results of the benchmark calculation of the ground-state energy of
  $^{40}Ca$ using the six-component truncated Urbana $U14$ interaction, the correlation
  functions shown in Fig. \ref{Fig2} \cite{fab02}, and harmonic oscillator
  mean field wave functions  with HO parameter $a=1.654$ $fm$(see Eq. (\ref{eqHOwf})). The results of the
  $\eta-$expansion obtained in this paper are compared
  with the FHNC/SOC results of Ref. \cite{fab02}. Notations are the same as in Table I. All quantities in $MeV$.}
    \begin{center}
      \begin{tabular}{l||c|c|c|c}
approach&$\langle V\rangle$&$\langle T\rangle$&$E$&$E/A$\\\hline
$\eta$ expansion, \textit{this paper}
                 & -1655.15 & 1425.90 & -229.25 & -5.73\\
\textit{FHNC/SOC}, Ref. \cite{fab02}
                 & -1891.60 & 1587.60 & -314.80 & -7.87\\\hline
      \end{tabular}
   \end{center}
\label{Table3}
\end{table}
%--------------------------------------------------------------------Table IV
%\newpage
\begin{table}[!hp]
  \caption{The contributions of the first six channels of the $U14$ interaction to
     the average potential energy $\langle V\rangle$ shown in Table. \ref{Table3}.
     Notations are the same as in Table I.
     All quantities in $MeV$.}
   \begin{center}
     \begin{tabular}{l||c|c|c|c|c|c|c}
approach&$\langle V_{c}\rangle$&$\langle V_{\sigma}\rangle$&$\langle V_{\tau}\rangle$
        &$\langle V_{\sigma\tau}\rangle$&$\langle V_{t}\rangle$&$\langle V_{t\tau}\rangle$
        &$V=\sum_{i} V_i$\\\hline
$\eta$-expansion, \textit{this paper}
            & -14.57 & 83.20 & 91.93  & -1353.45 & 11.61 & -473.87& -1655.15\\
\textit{FHNC/SOC}, Ref. \cite{fab02}
& -8.40 & 92.00 & 108.40 & -1549.20 & 11.60 & -565.60 & -1891.20\\\hline
      \end{tabular}
   \end{center}
\label{Table4}
\end{table}
%---------------------------------------------------------------------Table V
\newpage
\begin{table}[!hp]
   \caption{The results of the calculation of the ground-state
   energy and radius of \ossi{ }using the full \vuotto{ }interaction, the correlation functions
   shown in Fig. \ref{Fig3} \cite{fab01} and HO and SW  mean field wave functions.
   The value of the HO parameter is $a=2.0$ $fm$ and the parameters of the SW well are as
   follows: $V_o = 42.0$ $fm$, $R_o = 3.6$ $fm$ and $a_o = 0.55$ $fm$.
   The results of the $\eta$-expansion
   obtained in this paper are compared  with the FHNC/SOC results of Ref. \cite{fab01}.
   Notations are the same as Table I. $<r^2>$ is the rms radius. Energies in $MeV$, radii in $fm$.}
   \begin{center}
     \begin{tabular}{l||c|c|c|c|c|c}
Mean Field &Approach&$\langle V\rangle$  &$\langle T\rangle$&
                                                          $E$&$E/A$&${<r^2>}^{1/2}$\\\hline
HO&$\eta$-expansion, \textit{this paper}
                 & -420.39 & 350.39 & -67.54 & -4.40& 2.99\\
HO&\textit{FHNC/SOC}, Ref. \cite{fab01}
                 & -439.84 & 353.44 & -86.40 & -5.40& 3.03\\\hline
      \end{tabular}
   \end{center}
\label{Table5}
\end{table}
%---------------------------------------------------------------------Table VI
\begin{table}[!hp]
   \caption{The same as in Table \ref{Table5} but for Woods-Saxon mean field
   wave functions.}
    \begin{center}
      \begin{tabular}{l||c|c|c|c|c|c}
Mean Field &Approach&$\langle V\rangle$&$\langle
T\rangle$&$E$&$E/A$&${<r^2>}^{1/2}$\\\hline SW&$\eta$-expansion,
\textit{this paper} & -500.59 & 444.10 & -56.50 & -3.50& 2.64\\
 SW&\textit{FHNC/SOC} \cite{fab01}
                 & -519.68 & 428.16 & -91.52 & -5.72& 2.83\\\hline
      \end{tabular}
   \end{center}
\label{Table6}
\end{table}
%---------------------------------------------------------------------Table VII
\newpage
\begin{table}[!hp]
   \caption{The same as in Table \ref{Table5}, for $^{40}Ca$;
   the value of the HO parameter is $a=2.1$ $fm$.}
   \begin{center}
      \begin{tabular}{l||c|c|c|c|c|c}
Mean Field &Approach&$\langle V\rangle$  &$\langle T\rangle$&
                                                          $E$&$E/A$&${<r^2>}^{1/2}$\\\hline
HO&$\eta$-expansion, \textit{this paper}& -1320.22  & 1048.22 & -272.00 & -6.80& 3.72\\
HO&\textit{FHNC/SOC} \cite{fab01}
                 & -1521.20 & 1193.60 & -327.60 & -8.19& 3.65\\\hline
      \end{tabular}
   \end{center}
\label{Table7}
\end{table}
%---------------------------------------------------------------------Table VIII
\begin{table}[!hp]
   \caption{The same as in Table \ref{Table7} but for the Woods-Saxon well with parameters
    $V_o = 50.0$ $fm$, $R_o = 5.3$ $fm$, $a_o = 0.53$ $fm$.}
   \begin{center}
      \begin{tabular}{l||c|c|c|c|c|c}
Mean Field &Approach&$\langle V\rangle$&$\langle T\rangle$&$E$&$E/A$&${<r^2>}^{1/2}$\\\hline
SW&$\eta$-expansion, \textit{this paper} & -1293.96 & 1018.19 & -275.77 & -7.00 & 3.75\\
SW&\textit{FHNC/SOC} \cite{fab01}
                 & -1547.20 & 1215.20 & -332.00 & -8.3 & 3.66\\\hline
      \end{tabular}
   \end{center}
\label{Table8}
\end{table}
%---------------------------------------------------------------------Table IX
\newpage
\begin{table}[!htp]
  \caption{The values of the parameters appearing in the parametrization given by
  Eqs. \ref{o16corfuncA}-\ref{o16corfuncB}
   of the correlation
    functions of  $^{16}O$ shown in Fig. 3.}
    \begin{center}
      \begin{tabular}{l||c|c|c|c|c|c||c|c|c|c|c}\hline
n  & $A^{(n)}_1$ & $A^{(n)}_2$ & $A^{(n)}_3$ & $A^{(n)}_4$ & $A^{(n)}_5$ & $A^{(n)}_6$
                 & $B^{(n)}_1$ & $B^{(n)}_2$ & $B^{(n)}_3$ & $B^{(n)}_4$ & $B^{(n)}_5$\\\hline
1  & 1.0005 & 0.37314 & -1.1781 & 0. & 0. & 0. &1.0& 2.0 & 0. & 0. &0.\\\hline
2  & 0. & -0.1372 & 0.1916 & -0.0226 & -0.0141 & 0.& 5.0 & 3.5 & 1.0 & 0.13& 0.\\\hline
3  & 0. & -0.0795 & 0.1271 & -0.0121 & -0.0330 & 0.& 5.0 & 3.5 & 1.5 & 0.14& 0.\\\hline
4  & 0. & -0.3817 & 0.4863 & -0.0535 & -0.0424 & 0.& 4.5 & 3.7 & 1.6 & 0.15& 0.\\\hline
5  & 0. & 0.0114  & 0.0527 & -0.0702 & 0.0064  & 0.& 0.8 & 1.5 & 1.5 & 1.5 & 0.\\\hline
6  & 0. & -0.1776 & -0.0054& -0.0237 & -0.00006& 0.& 1.7 & 1.0 & 1.3 & 0.01& 0.\\\hline
      \end{tabular}
   \end{center}
\label{Table9}
\end{table}

%---------------------------------------------------------------------Table X
\begin{table}[!hbp]
  \caption{The same as in Table \ref{Table9} for  the correlation functions of
  $^{40}Ca$ shown in Fig. 4.}
    \begin{center}
      \begin{tabular}{l||c|c|c|c|c|c||c|c|c|c|c}\hline
n  & $A^{(n)}_1$ & $A^{(n)}_2$ & $A^{(n)}_3$ & $A^{(n)}_4$ & $A^{(n)}_5$ & $A^{(n)}_6$
               & $B^{(n)}_1$ & $B^{(n)}_2$ & $B^{(n)}_3$ & $B^{(n)}_4$& $B^{(n)}_5$\\\hline
1  & 1.00039 & 0.7576 & -1.6015 & 0. & 0. & 0. & 3.9 & 2.9 & 0. & 0. & 0.\\\hline
2  & 0. & -0.0573 & 0.0965 & -0.0156 & -0.0145 & 0. & 7.0 & 3.5 & 0.8 & 0.2 & 0.\\\hline
3  & 0. & -0.0207 & 0.0474 & -0.0019 & -0.0279 & 0. & 8.5 & 3.5 & 2.0 & 0.215 & 0.\\\hline
4  & 0. & -0.0290 & 0.0906 & -0.0237 & -0.0356 & 0. & 9.4 & 3.0 & 1.0 & 0.22 & 0.\\\hline
5  & 0. & 0.0165 & 0.0061 & 0.0009 & -0.0188 & -0.0041 & 1.0 & 3.0 & 0.3 & 1.3 & 4.5\\\hline
6  & 0. & -0.1342 & 0.00013 & -0.0368 & 0.00044 & -0.00069 & 1.55 & 0.02 & 1.4 & 0.1 & 0.1
                       \\\hline
      \end{tabular}
   \end{center}
\label{Table10}
\end{table}
%---------------------------------------------------------------------Table XI
\newpage
\begin{table}[!hp]
   \caption{The value of $C^{(p,q)}_D(\Vec{r}_{12},\Vec{r}_{1^\prime 2})$ and
   $C^{(p,q)}_E(\Vec{r}_{12},\Vec{r}_{1^\prime 2})$ defined in Appendix C. The order of the operator
   ${p,q}={1,2,\ldots,6}$ is the same as Table \ref{Table2}. Here
   $\langle S_{12}S_{1'2}\rangle$ is defined as
   $\langle S_{12}S_{1'2}\rangle=12(3(\Vec{\hat{r}}_{12}\cdot\Vec{\hat{r}}_{1'2})^2-1)$,
   with the definition of $\Vec{\hat{r}}=\Vec{r}/r$.}

   \begin{center}
     \begin{tabular}{|r||c|c|c|c|c|c|}
\hline
Operator&  &   &  &   &   & \\
p/q   & \,\,\,\,$1$ & \,\,\,\,$2$  & \,\,\,\,$3$ &  \,\,\,\,$4$ &  \,\,\,\,$5$ &  \,\,\,\,$6$ \\
\hline
$1$ \,\,\,\,D & 16  &  0   &  0  &   0  &  0   &  0   \\
    E &  4  &  12  &  12 &  36  &  0   &  0   \\ \hline
$2$ \,\,\,\,D &     &  48  &  0  &   0  &  0   &  0   \\
    E &     & -12  &  36 &  -36 &  0   &  0   \\ \hline
$3$ \,\,\,\,D &     &      &  48 &   0  &  0   &  0   \\
    E &     &      & -12 &  -36 &  0   &  0   \\ \hline
$4$ \,\,\,\,D &     &      &     &  144 &  0   &  0   \\
    E &     &      &     &   36 &  0   &  0   \\ \hline
$5$ \,\,\,\,D &     &      &     &      & $4\langle S_{12}S_{1'2}\rangle$ &  0   \\
    E &     &      &     &      & $2\langle S_{12}S_{1'2}\rangle$ & $6\langle S_{12}S_{1'2}\rangle$ \\ \hline
$6$ \,\,\,\,D &     &      &     &      &      &  $12\langle S_{12}S_{1'2}\rangle$   \\
    E &     &      &     &      &      &  $-6\langle S_{12}S_{1'2}\rangle$    \\ \hline

      \end{tabular}
   \end{center}
\label{TableMatrix}
\end{table}

%%%%%%%%%%%%%%%%%%%%%%%%%%%%%%%%%%%%%%%%%%%%%%%%%%%%%%%%%%%%%%%%%%%%%%%%%%%%%%---------------------------------------------------------------------- FIG I
\newpage
%---------------------------------------------------------------------- FIG II
%---------------------------------------------------------------------- FIG III
\newpage
%---------------------------------------------------------------------- FIG IV
%---------------------------------------------------------------------- FIG V
\newpage
%---------------------------------------------------------------------- FIG VI
\newpage
%---------------------------------------------------------------------- FIG VII
\newpage
%---------------------------------------------------------------------- FIG VIII
\newpage
%---------------------------------------------------------------------- FIG IX
\newpage
%---------------------------------------------------------------------- FIG X
\newpage
%---------------------------------------------------------------------- FIG XI
\newpage
%---------------------------------------------------------------------- FIG XII
\newpage
%---------------------------------------------------------------------- FIG XIII
\newpage

%---------------------------------------------------------------------- FIG XIV
%\newpage
%---------------------------------------------------------------------- FIG XV
\newpage
%---------------------------------------------------------------------- FIG XVI
\newpage
%---------------------------------------------------------------------- FIG XVII
\newpage
%---------------------------------------------------------------------- FIG XVIII
%\newpage
%------------------------------------------------------------------------------
\end{document}